\documentclass[]{geophysics}

\usepackage[]{algorithm2e}
\usepackage{algorithmic}

\usepackage{amsmath}    
\usepackage{graphicx}   
\usepackage{verbatim}   
\usepackage{color}      
\usepackage{hyperref}   
\newcommand{\Vect}[1]{\rm Vec(#1)}
\newcommand{\Tr}[1]{\rm tr(#1)}
\newcommand{\Sign}[1]{\rm sign(#1)}
\newcommand{\argmin}{\arg\!\min}
\newcommand{\MVN}{{\rm MVN}}

\newcommand{\gmat}[1]{{\bf \uppercase{#1}}}
\newcommand{\gvec}[1]{{\bf \lowercase{#1}}}

\DeclareGraphicsExtensions{.pdf,.png,.jpg}

\begin{document}

\title{A novel prestack sparse azimuthal AVO inversion}
\author{B. G. Lasscock, T. A. Sansal}

\begin{abstract}

  In this paper we demonstrate a new algorithm for sparse prestack
  azimuthal AVO inversion. A novel Euclidean prior model is developed
  to at once respect sparseness in the layered earth and smoothness in
  the model of reflectivity. Recognizing that methods of artificial
  intelligence and Bayesian computation are finding an every
  increasing role in augmenting the process of interpretation and
  analysis of geophysical data, we derive a generalized matrix-variate
  model of reflectivity in terms of orthogonal basis functions,
  subject to sparse constraints. This supports a direct application of
  machine learning methods, in a way that can be mapped back onto the
  physical principles known to govern reflection seismology. As a
  demonstration we present an application of these methods to the
  Marcellus shale.  Attributes extracted using the azimuthal inversion
  are clustered using an unsupervised learning algorithm. Interpretation
  of the clusters is performed in the context of the Ruger model of
  azimuthal AVO.
  
\end{abstract}

\maketitle

\section{Introduction}

The usual method for approaching AVO inversion is to presume a
particular physical model of reflectivity, the two- or three-term
Shuey or Aki-Richards models for instance \cite[]{Castagna:93}, and
regress this against collections of migrated data binned to create
a regular set of angle gathers. For extension to an azimuthal
inversion we would choose a particular two- or three- term version of
Ruger's model \cite[]{Cooke:16}, regressed against a set of
azimuthally sectored angle gathers. The utility of these methods is to
provide attribute maps for inversion of elastic properties or just to
aid interpretation. Methods of artificial intelligence (A.I.) in the
form of ``multi-variate statistics'' are commonly applied to
accelerate the interpretation of seismic attributes. With the
principle application of machine learning to identify or discriminate
areas in the subsurface based on metrics of similarity. Similarly,
applications of A.I. to well-log analysis are already well established
in the geophysical literature, ~\cite[]{Hall:17} and
\cite[]{Schlanser:16} are recent examples supervised and unsupervised
learning applications.  The key feature in the application of A.I. to
well-log analysis is that these systems learn relationships in a
non-parametric data driven fashion.
And as such the methods are portable as we add additional well
information. On the other-hand, seismic inversion of AVO attributes
generally require {\it a priori} selection of a physical model and
other interpretive methods in order to arrive at a volumes of
attributes.

%
%
%
Learning algorithms do not necessarily require physically
motivated characteristics of the subsurface to metric
similarity. For the purpose of cataloging data we only need to
ensure that the parameter space of the attributes used by learning
algorithms remains invariant (as possible) between analyses, which
provides a challenge.
Since seismic data cannot identify a unique model of reflectivity,
some additional prior knowledge is required to identify the
problem. Further, while the AVO intercept and gradient are attributes
that we might use classify points on a horizon in the subsurface,
extensions to an azimuthal inversion at become explicitly dependent on
the method used to resolve the ambiguity in the orientation of the
fractures. While higher order approximations to the Zoepritz equations
necessary for analyzing far offset data typically introduce degrees of
freedom that tend to be co-linear at smaller takeoff angles, making
the effective number of degrees of freedom in the parameter space
dependent on the acquisition.  Other details such as balancing fold
within the angle- and sectored gathers introduces another dimension of
variability in the meaning of the extract attributes. And finally,
consider a scenario where there is also variability in the data from
an exogenous source, this may result from processing methods,
attenuation or other effects. In this case, does a two- or three- term
AVO model have enough degrees of freedom to properly capture this
variability? If not, then we maybe leaving behind as ``noise''
coherent structure that machine learning algorithm might otherwise
find relevant.

The direction of this paper will be to start by formulating the
azimuthal AVO (AVO(z)) inversion in a general framework, minimizing
the project specific characteristics of the process. The model we
invert is independent of the specific physical model of reflectivity,
other than viewing the earth as a convolution of reflectivity
convolved with a wavelet. The inversion is constructed in such a way
as to support direct application to common-offset gathers or offset
vector tiles. For many applications we can assume that by construction
the seismic survey, while not necessarily regularly sampled in this
domain, will generally be uniform.

The workflow we present here is to first determine what independent
attributes are required to explain the variability in the observed
seismic data. 
%
Attributes are estimated that best describe the shape of the
reflectivity surface as a function of offset and azimuth. These
attributes themselves lie in a higher dimensional parameter space. We
then use machine learning in this parameter space to determine how, or
if, these attributes exhibit clustering in this parameter space. If
they do exhibit clustering, then we may classify each point in the
subsurface by associating it with a particular cluster. Interpretation
of the clusters is the final step in the workflow. In this paper,
interpretation is done by projecting the problem onto a particular
physical model, where we find that we are able to discriminate areas
that exhibit fracturing (or otherwise) in an application to the
Marcellus shale.


The layout of the paper is as follows, first we provide a mathematical
description of the statistical properties of the model and detail our
Bayesian priors. Next we detail the method used for interpretation
within our workflow by expanding a two term version of Ruger's model
\cite[]{Ruger:97} in terms of the estimated attributes. As a
demonstration, we provide an analysis of a synthetic data, and then a
real dataset taken from the Marcellus shale.  We see how machine
learning is used to augment the interpretation of the AVO(z)
inversion. And finally we present our conclusions.  Mathematical
details of the optimization of the prestack sparse azimuthal
inversion with sparse grouped constraints are outlined in the Appendices.

\section{Geological Setting}
\label{sec:geo_setting}

\begin{figure}
  \begin{center}
    \includegraphics[height=8.5cm]{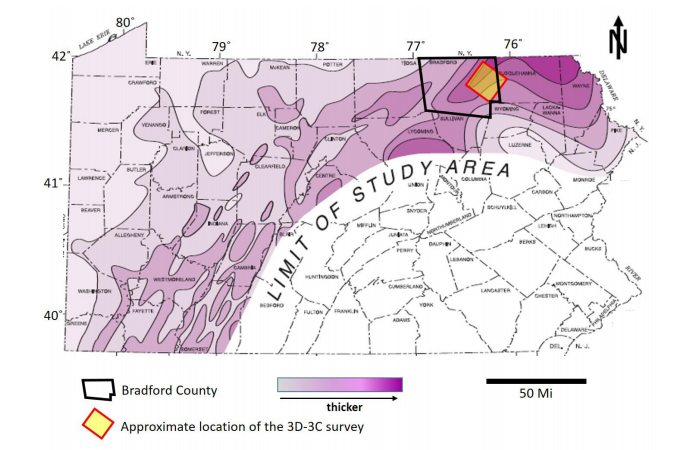}
    \end{center}
  \caption{\label{fig:shale_thickness} A map of Marcellus shale thickness in
    Pennsylvania overlain by the Bradford County and the approximate
    location of the survey outline. Figure reproduced from the
    masters thesis~\cite{Pelayo:16}.}
\end{figure}

Figure~\ref{fig:shale_thickness} Illustrates the location of a 3D-3C
survey acquired by Geokinetics Inc., and Geophysical Pursuit Inc.
using dynamite sources covering 12.3 ${\rm mi}^2$
\cite[]{Chaveste:13}.  A subset of the processed PP (P-wave to P-wave)
component is used. The data is processed into offset vector tiles
(OVTs) for azimuthal velocity and AVO analysis. Raw OVT gathers
computed by the pre-stack time migration were later corrected for time
shifts caused by HTI anisotropy using elliptical inversion. The
results were conditioned by a pass of residual trim statics to further
flatten the gathers. After leveling, well recognized random noise
elimination workflows were applied to produce the optimal input to the
sparse AVAZ inversion.  Data examples revealed in the study are
recorded within the Allegheny Plateau Province of the Appalachian
Basin in Bradford County, Pennsylvania targeting the middle Devonian
Marcellus shale. Numerous organic rich shales, including the
Marcellus, were largely deposited in a foreland basin setting formed
by three major orogenic events. Collision of the North American plate
(Laurentia) and the eastern oceanic crust (Avalonia) transformed the
passive margin into a foreland basin. The Acadian orogeny occured in
the Middle Devonian and triggered flexural deformation. Between mid to
late Devonian, the basin was confined by the developing Acadian
mountains, Cincinnati arch, and the Old Red Sandstone continent
forming a nearly bounded epicontinental sea \cite[]{Wang:13}.  The
Marcellus is believed to have been deposited over a 2 m.y. duration in
a deep and anorexic water environment, which helped accumulate
hydrocarbons with minimal breakdown \cite[]{Wang:13, Lash:11,
  Koesoemadinata:11}.  The Marcellus is a part of the Hamilton group
(7-8 m.y. timeframe) and can be found at the base of the set confined
by Stafford limestone above and Onondaga limestone below. The
organically rich upper and lower parts (also known as “Hot” Marcellus)
of the black shale are divided by the Cherry Valley limestone. The
complete group is decoupled from the deeper structures by the Silurian
Salina salt formation. This produces shorter wavelength folds than
deeper formations \cite[]{Chaveste:13}. The average gross thickness of
the Marcellus is approximated to be around 80 ft. The organic rich
average thickness (where TOC $> 6.5\%$) is 34 ft. TOC can reach up to
20\% \cite[]{Wang:13}.  Engelder et al.\cite[]{Engelder:09} shows that
Marcellus shale exhibits two regional joint sets referred to as J1 and
J2 sets that make up the natural fracturing network. J1 joints are
nearly parallel to the maximum horizontal stress direction and are
closely spaced. The younger J2 joints are orthogonal to the J1
joints. Parallelism of J1 to maximum horizontal stress field (N75E
direction), which is induced by tectonic stress, is concluded to be a
coincidence. The joints are understood to be formed due to the
presence of abnormal pressure during maturation following the
deposition \cite[]{Engelder:08, Engelder:09}. The J1 and J2 joint sets
make it favorable to drill horizontal wells for hydrocarbon
production. While NNW/SSE (perpendicular to maximum horizontal stress
direction) directional drilling drains the J1 sets, later hydraulic
stimulation can cross-cut the J2 sets and deliver hydrocarbons
\cite[]{Engelder:09, Enomoto:11}.  High spatial variability of
mineralogy, fracturing, and the limits of seismic resolution create a
big challenge in reservoir characterization using conventional narrow
azimuth data. Examining the HTI time anomalies and joint understanding
of amplitude variation with azimuth with this afresh technology can
assist in classification of the highly desirable natural fracture
networks and support determining sweet spots for hydrocarbon
exploration.

\section{Formalism}
\label{sec:formalism}

\subsection{General notation}

Throughout this document, we are going to indicate variables that are
matrices in capitals $\gmat{Y}$, and vectors using an the lowercase bold
$\gvec{y}$. Elements of a matrices and vectors will be labeled with
subscript notation; e.g. $y_{ij}$ and $y_{i}$ for matrices and vectors
respectively. The upper case super/subscript notation $\gmat{Y}^{l}$ and
$\gmat{Y}_{l}$ indicates the rows and columns of the matrix respectively.
The result being row and column vectors; the notation $\gmat{Y}^{T}_{l}$ is
meant to indicate that we are (in this case) taking the column of the
matrix as a column vector and then transposing it to a row vector. As
a special case, the symbol $\gmat{\mathcal{I}}_{n}$, will represent the identity
matrix, where the subscript indicates the rank of the
matrix. Otherwise variables are considered to be scalars (or functions
where appropriate). In terms of operators, $\Sign{\gmat{B}^{l}}$ is a vector
whose elements are $\Sign{b_{li}}$, and the derivative
$\frac{\delta}{\delta \gmat{B}^{l}}$ should be interpreted as a gradient (a
vector) whose elements are $\frac{\delta}{\delta b_{li}}$.

For the purpose of efficient computation, it is useful to use the
Kronecker product and vectorization operators to account for the
layout of the model.  The Kronecker product $\gmat{A}\otimes \gmat{B}$, for a
$(m\times n)$ matrix $\gmat{A}$, and $(p\times q)$ matrix $\gmat{B}$, for example, is:
\[ \gmat{A}\otimes \gmat{B}  = \left[ \begin{array}{cccc}
  a_{11}\gmat{B} & . & . & a_{1n} \gmat{B} \\
   . & . &   &    \\
   . &   & .  &    \\
  a_{m1}\gmat{B} &   &  & a_{mn} \gmat{B} \\
\end{array} \right]\ .\]
The vectorization operator acting on a matrix, simply stacks the columns of
matrix into a vector:
\[ \Vect{\gmat{A}}  = \left( \begin{array}{c}
  a_{11} \\
  a_{21} \\  
  . \\
  . \\
  a_{nm} \\
\end{array} \right)\ .\]
A collection of useful identities associated with operations on matrices \gmat{A}, \gmat{B}, and \gmat{C} are:
\begin{eqnarray}
  &&{\rm Inverse: } (\gmat{A}\otimes \gmat{B})^{-1} = (\gmat{A}^{-1}\otimes \gmat{B}^{-1}) \cr
  &&{\rm MatMult: }(\gmat{A}\otimes \gmat{B}) (\gmat{C}\otimes \gmat{D}) = (\gmat{AC}\otimes \gmat{BD})\cr  
  &&{\rm MatVec: } (\gmat{C}^{T}\otimes \gmat{A})\Vect{\gmat{B}} = \Vect{\gmat{ABC}} \cr
  &&{\rm Vec: } \Vect{\gmat{AB}} = (\gmat{B}^{T}\otimes \gmat{\mathcal{I}}_{n})\Vect{\gmat{A}}\ ,
\end{eqnarray}
for the final identity, $n$ is the number of rows of the matrix $\gmat{A}$.

To denote statistical models, we use the notation $p(\gmat{A},\gmat{B})$ to mean the
joint probability of \gmat{A} and \gmat{B}, and $p(\gmat{A}|\gmat{B})$ to denote the conditional
probability. Bayes theorem would be expressed as:
\begin{eqnarray}
p(\gmat{A},\gmat{B}) &=& p(\gmat{A}|\gmat{B})p(\gmat{B})\ .
\end{eqnarray}

\subsection{Observational model}
\label{ssec:liklihood}

The input to the inversion is a seismic wavelet and a time migrated
seismic data, organized by sectored angle and azimuth gathers or into
offset vector tiles.  We are going to structure the input seismic data
as a matrix, with temporal and spatial (offset and azimuth)
dimensions. We do not need to assume that the observations are
regularly sampled by offset (or angle), only that the sampling in the
offset and azimuthal domain is uniform.

Define a matrix $\gmat{Y}$ of observations, the columns of which contain a
trace of data at a particular coordinate in the offset/azimuth
domain. The convolutional model of the earth expresses an observed
seismic trace as a convolution of a wavelet with a sequence of
reflectivities. The reflectivity is a function both of temporal and
spatial coordinates of the trace, we will represent the reflectivity as a
matrix $\gmat{R}$, which has the same dimensions as $\gmat{Y}$:
\begin{eqnarray}
\gmat{Y}^{i}_{t} &=& (w * \gmat{R}^{i})_{t} + \gvec{\epsilon}_{t}\ ,
\end{eqnarray}
here $w$ represents the seismic wavelet and $\epsilon$ represents
idiosyncratic noise (considered exogenous to the model), both are
functions of time.

Continuing with the matrix notation, since the time sampling of the
data is discrete, we can represent the convolution as a matrix-vector
product. Suppose we have $T$ time samples and $N$ combinations of
offset and azimuth in the gather, and that we use the same seismic
wavelet for each coordinate. Then the observational model can be
represented as a matrix vector product:
\begin{eqnarray}
\label{eq:obs}
\Vect{\gmat{Y}} &=& (\mathcal{I}_{N}\otimes \gmat{W})\Vect{\gmat{R}} + \Vect{\gmat{E}}\ ,
\end{eqnarray}
where $\gmat{W}$ is a convolution matrix build from the seismic wavelet
(which is circulant) dimension ($T\times T$). Matrices $\gmat{Y}$, $\gmat{R}$ and
$\gmat{E}$ contain the observations, reflectivity and noise respectively, each matrix
has dimensions ($T\times N$). 

The reflectivity at each point in time is very generally a function of
offset and azimuth.  We can represent any function on this domain as
an expansion in some complete polynomial basis (or stacks),
\begin{eqnarray}
  \gmat{R} &=& \gmat{B} \gmat{A}^{T}\ , 
\end{eqnarray}
here $\gmat{B}$ is a $(T\times r)$ matrix and $\gmat{A}$ is a $(N\times r)$. The
magnitude of the coefficients in the columns of the matrix $\gmat{B}$
represent the weightings of the stacks. A constraint on the
normalization of $\gmat{A}$ become necessary for identifying the model; a
convenient normalization is such that $\gmat{A}$ is orthogonal.
Re-factoring our observational model to a more convenient form:
\begin{eqnarray}
  \label{eqn:matvec}
  \Vect{\gmat{Y}} &=& (\mathcal{I}_{N}\otimes \gmat{W})\Vect{\gmat{B} \gmat{A}^{T}} + \Vect{\gmat{E}}\cr
           &=& (\mathcal{I}_{N}\otimes \gmat{W})(\gmat{A}\otimes \mathcal{I}_{T})\Vect{\gmat{B}} + \Vect{\gmat{E}}\cr
           &=& (\gmat{A} \otimes \gmat{W})\Vect{\gmat{B}} + \Vect{\gmat{E}} \ .
\end{eqnarray}
With the useful identities from the previous section, we can show that this is equivalent
to a matrix equation of reflectivity convolved with a wavelet plus some idiosyncratic noise:
\begin{eqnarray}
  \label{eqn:not_linear}
  \gmat{Y}        &=& \gmat{W}\gmat{B} \gmat{A}^{T} + \gmat{E}\ .
\end{eqnarray}

\subsection{Statistical model and rank reduction}

For the purpose of optimization, the form of Eq.~\ref{eqn:not_linear}
is not convenient because it is not linear in the set of parameters
$\gmat{B}$. To derive an equivalent linear problem, we have to first
decided what the structure of the background noise is going to be.
As an approximation, we will assume that the noise is normally
distributed, and that the noise-model can be factorized in terms of
the spatial ($\gmat{Q}$) and temporal ($\gmat{\Sigma}$) degrees of freedom
independently,
\begin{eqnarray}
  \gmat{E} \sim \gmat{\Sigma} \otimes \gmat{Q}, 
\end{eqnarray}
In this case the likelihood function we would optimize for with
respect to the matrix of coefficient $\gmat{B}$ is matrix-variate
normal (MVN)~\cite[]{Gupta:09},
\begin{eqnarray}
  \mathcal{L}(\gmat{Y}|\gmat{B}) = MVN(\gmat{Y}; \gmat{B}\gmat{A}, \gmat{\Sigma}, \gmat{Q})\ .
\end{eqnarray}
To make the problem more tractable computationally, we will make the
usual assumptions that the noise model can be represented as a scaled
identity, $\gmat{\Sigma} = \sigma^{2}\mathcal{I}_{T}$ and
$\gmat{Q}=\mathcal{I}_{N}$.  This model for $\gmat{\Sigma}$ means that we expect the
noise is constant in time, and not correlated temporally. The model
covariance $\gmat{Q}$ means that we do not expect that the noise should be
correlated between traces within a gather.

Now consider that spatially, reciprocity means that solution for an
AVO analysis must only be non-zero for even basis functions. Therefore
we know that the reflectivity matrix $\gmat{R}$ must have a reduced
rank. There are also physical reasons to suppose that the solution
space should have a low dimension. For an application to
unconventional resources, \cite[]{Glinsky:15} uses the floating grain
model to decompose the AVO response in terms of its independent
degrees of freedom. They find that only two- or three-independent
pieces of information are likely resolvable for most applications.
Practically this means that we can truncate the expansion at some
order $r$ in our polynomial basis of choice. This truncation is
equivalent to use setting the appropriate coefficient on the matrix
$\gmat{B}$ to zero. Following the approach of Chen
et~al.\cite[]{Chen:12}, consider a complete set of orthonormal basis
vectors packed into the matrix \gmat{A},
\begin{eqnarray*}
\gmat{A} = [\gmat{A}_{\bot}, \gmat{A}_{\parallel}]\ ,
\end{eqnarray*}
where $\gmat{A}_{\bot}$ is a (N$\times$r) matrix and $\gmat{A}_{\parallel}$ is a
(N$\times$(N-r)) matrix. To truncate the expansion, the coefficients
of the matrix $\gmat{A}_{\parallel}$ are set to zero.
Taking advantage of the orthogonality of the basis functions, and our
assumptions about the error model, the likelihood function becomes: 
\begin{eqnarray}
  \label{eq:bot}
  \mathcal{L}(\gmat{Y}|\gmat{B}) &=& \MVN(\gmat{Y}; \gmat{W}[\gmat{B}_{\bot},0][\gmat{A}_{\bot},\gmat{A}_{\parallel}]^{T},
  \sigma^{2}\mathcal{I}_{T}, \mathcal{I}_{N})\cr
  &=& \MVN(\gmat{Y}\gmat{A}_{\bot}; \gmat{W}\gmat{B}_{\bot}, \sigma^{2}\mathcal{I}_{T}, \
              I_{r})MVN(\gmat{Y}\gmat{A}_{\parallel}; 0, \sigma^{2}\mathcal{I}_{T}, \mathcal{I}_{N-r})\ .
\end{eqnarray}
The above is just statement of conditional probability of the
matrix-variate normal. The second term in this equation is just a
scalar.  In the case where our input data is not regularly sampled, or
where we what to regress on the more common ${1, \sin^{2}, tan^{2}}$
basis, the approach is still applicable, but we need to use a
QR-decomposition and solve an analogous preconditioned problem, this
procedure is detailed in Appendix A.
Dropping the ${\bot}$ subscripts from Eq.~\eqref{eq:bot}, we arrive at a simplified version
of our likelihood function:
\begin{eqnarray}
  \label{eq:master}
  \mathcal{L}(\gmat{Y}|\gmat{B}) &\propto& \MVN(\gmat{Y}\gmat{A}; \gmat{W}\gmat{B}, \sigma^{2}\mathcal{I}_{T}, I_{r}).
\end{eqnarray}
Optimizing this linear function is equivalent to solving
Eq.~\ref{eqn:not_linear}, given the assumptions discussed above.
An optimization algorithm for this equation subject to sparse penalties
is derived in the Appendices.

\subsection{Prior model - sparsity and smoothness}

Temporally, our {\it a priori} expectation is that the earth response
should be sparse and heavy tailed. This is consistent with an
assumption of a layered earth, with relatively many bright reflectors
(compared to Gaussian assumption). This is a common assumption made
for post-stack sparse spike inversion, see \cite[]{Zhang:07} as an
example.  Spatially; common models of amplitude variation by offset
and azimuth, such as a Shuey and Ruger models are smooth. Distortion
due to effects of NMO-stretch or attenuation should also be
smooth. And so our prior model should reflect the sparse layered
earth, and otherwise smooth amplitude variation by offset.

Post-stack, the reflectivity (as a function of offset) is projected
onto constant stack weights to form a time-series. The L1-norm of this
time-series reflects the prior assumption that the earth is
sparse. Generalizing this to the pre- (or multi-) stack example, the
negative log-prior (up to a constant in $\gmat{B}$) is:
\begin{eqnarray}
  \label{eq:naive}
  p(\gmat{B}) &=& \lambda \Vert \Vect{\gmat{B}} \Vert_{1}\ ,
\end{eqnarray}
where $\lambda$ is the rate parameter of the Laplace
distribution. This kind of prior assumption induces a sparse
solution\cite[]{Sassen:13} and \cite[]{Sassen:15}, however, except in
the special case of a post-stack inversion, we cannot directly
interpret this in terms of the sparse layered earth. Rather, this
prior makes a statement regarding the overall complexity of the
solution.

Instead we introduce a new prior model for sparse inversion of seismic data
based\cite[]{Chen:12},
\begin{eqnarray}
  \label{eq:grouped}
  p(\gmat{B}) &=& \sum_{t=1}^{T} \lambda_{t} \Vert \gmat{B}^{t} \Vert_{2} \ ,
\end{eqnarray}
where $\lambda_{t}$ is now a collection of rate parameters, on for
each point in time, and $\Vert \cdot \Vert_{2}$ is the Euclidean
norm. Following the statistical literature\cite[]{Friedman:10}, we can
view the parameterization of the AVO(z) response at each time as a
group. This prior induces sparsity in terms of the groups, the
inversion will then be sparse in terms of which groups are active
(in-active groups are thresholded uniformly to zero).
Where, in our application, an active group at a point in time
represents the location of a reflector.
Otherwise, we are not enforcing a sparse penalty on the complexity of
the AVO(z) response within each group. Therefore, this prior model
represents the belief that the AVO(z) response is smooth, but the
earth is sparse and layered.
Note, there is a crucial difference in behavior between the canonical
L1 and sparse grouped priors for the specific application to an
azimuthal inversion. Consider that the change in reflectivity is a
function of azimuth measured relative some axis of symmetry that is
{\it a priori} unknown. The regression problem must be applied to a
set of azimuthally varying functions, where azimuth is defined in some
arbitrary orientation, with respect to true or grid North for example.
Since this orientation is arbitrary, the invariances of the prior
model subject to a rotation of the coordinate system is important.
The analysis of Chen\cite[]{Chen:12} shows that the location of the
reflectors in time, generated by the sparse grouped prior model
Eq.~\ref{eq:grouped} are guaranteed to be invariant under a rotation
of the coordinate system.  However, the usual L1-prior applied to the
azimuthal inversion is not.

Whilst both the noise model and importance of the layer based
prior may be functions of time, calibration of this expansive
parameter space is beyond the scope of this paper. For the remainder
of this article we will deal with the special case where both are
constant in time. Since our analysis will center around a limited time
window around the Marcellus shale, this is a reasonable
simplification.  Combining our prior model with the likelihood
function Eq.~\eqref{eq:master}, the negative log-posterior probability
(up to an additive constant in $\gmat{B}$) we seek to minimize is the
matrix-variate function:
\begin{eqnarray}
  \label{eq:posterior}
        {\rm log}(p(\gmat{B}|\gmat{Y})) &=& \frac{1}{2\sigma^{2}}
        \Vert \gmat{Y}\gmat{A}^{T} - \gmat{W}\gmat{B} \Vert^{2}_{2} +
        \lambda \Vert \Vect{\gmat{B}} \Vert_{1} +\
        \lambda_{g} \sum_{t=1}^{T} \Vert \gmat{B}^{t} \Vert_{2}\ .
\end{eqnarray}
The method for optimizing this model is detailed in
Appendix B. Note that we are detailing a particular choice for
the grouped constraint, where we group by row. However, the optimization
applies equally where the groups span for multiple rows of $\gmat{B}$.

\section{Method}
\label{sec:method}

The Appendices detail a very general model for inverting for the
azimuthally varying AVO response.  The inputs to the model are a
wavelet, extracted at a well using a 25-degree stack and a HTI
corrected set of time migrated offset vector tile (OVT) gathers.

The sparse inversion is performed on a time interval between the Tully
and Onondaga limestone. We linearly interpolate a set of hand picked
horizons across the survey. The inverted attributes are a sequence of
sparse spikes, which do not necessarily lie exactly along the
interpolated horizon. We therefore perform a search for the nearest
peak (or trough depending on the horizon) within 6ms of the horizon
picked at each inline and associate that with the spike. Since the
spikes themselves have a finite bandwidth, a secant area
method\cite[]{Glinsky:16} is used to integrate the spikes.


The region that is of particular commercial interest is the so called
Hot Marcellus, which is a thin layer of high total organic content
(TOC) shale which sits atop the Onondaga limestone. We will provide
interpretation of the inversion for this horizon. To classify each
point on the horizon using the estimated attributes, we apply a
unsupervised clustering Expectation Maximization (EM) algorithm.  The
EM-algorithm takes as input the expected number of clusters, it then
fits a multivariate Gaussian density to describe the distribution of
points within each cluster. The EM-algorithm is similar to the K-means
classifier commonly used in seismic attribute analysis, except that it
clusters are hyper-ellipsoidal rather than strictly hyper-spherical.
Given the elliptical relationships commonly seen cross-plotting
rock-physics relationships, the EM-classifier is more general and more
appropriate for this kind of attribute analysis.

We do not know the number of clusters {\it a priori},
our method for calibrating this is to sequentially add more clusters
to the model, stopping once we found a parsimonious description of the
horizon.
Assuming the elastic properties of the shale and limestone are fairly
stationary spatially, we expect that there should be at least two
classes of material, which we may expect to interpret as fractured or
non-fractured. Since we do not have sufficient well measurements to
allow us to characterize the clusters directly, we need to appeal to a
particular model for the AVO response for interpretation. Following
Tylasning~\cite[]{Cooke:16}, we will interpret the reflectivity in
terms of Ruger's two-term model\cite[]{Ruger:97}. Assuming Ruger's
model, we derive an estimate of the anisotropic gradient, and fracture
orientation, up to a $90^{\circ}$ ambiguity.
%
In a recent study by~\cite[]{Schlanser:16}, the EM-algorithm was
applied to well-log data for the purpose of classifying litho-facies.
This study identifies four classes of shale, and a carbonate in our
area of interest.  The results of this analysis are used to provide a
comparison with the two estimates of isotropic gradient predicted by
the Ruger model. The solution that is consistent with the regional
data resolves the $90^{\circ}$ ambiguity.
Note, the inversion will be performed with sufficient degrees of
freedom to allow us to interpret a higher order model, however we will
leave that as future work.



\subsection{Interpretation}
\label{ssec:ruger_param}

Consider an inversion of data which at each point in time has spatial
degrees of freedom of offset and azimuth. For example where the input
data is a set of offset vector tiles (OVT). The domain of the problem
is an azimuths ($\phi$) between 0-180-degrees (reciprocity makes the
backward angles redundant) and range of offsets ($x$) normalized to
the interval $[0,1]$.

%

We normalize measured offsets within the gather to on the interval
[0,1], azimuth is measured on the interval [0,$\pi$]. Reciprocity
demands that the reflectivity is symmetric under reflections, so we can
imagine that the domain of the problem by offset is [-1,1], but that
the solution should be symmetric, which we enforce. The basis of
spherical harmonic function $Y_{lm}(x,\phi)$ is complete on this
domain. Therefore very generally, any smooth function of the
reflectivity can be represented by this expansion:
\begin{eqnarray}
  r(x,\phi) &=& \sum_{lm} a_{lm} Y_{lm}(x,\phi) \ .
\end{eqnarray}
Here the coefficients $a_{lm}$ would be the most general set of attributes,
however, for computational efficiency we appeal to physical modeling
to reduce the dimension of this parameter space. 

\begin{figure}
  \begin{center}
    \includegraphics[height=5.5cm]{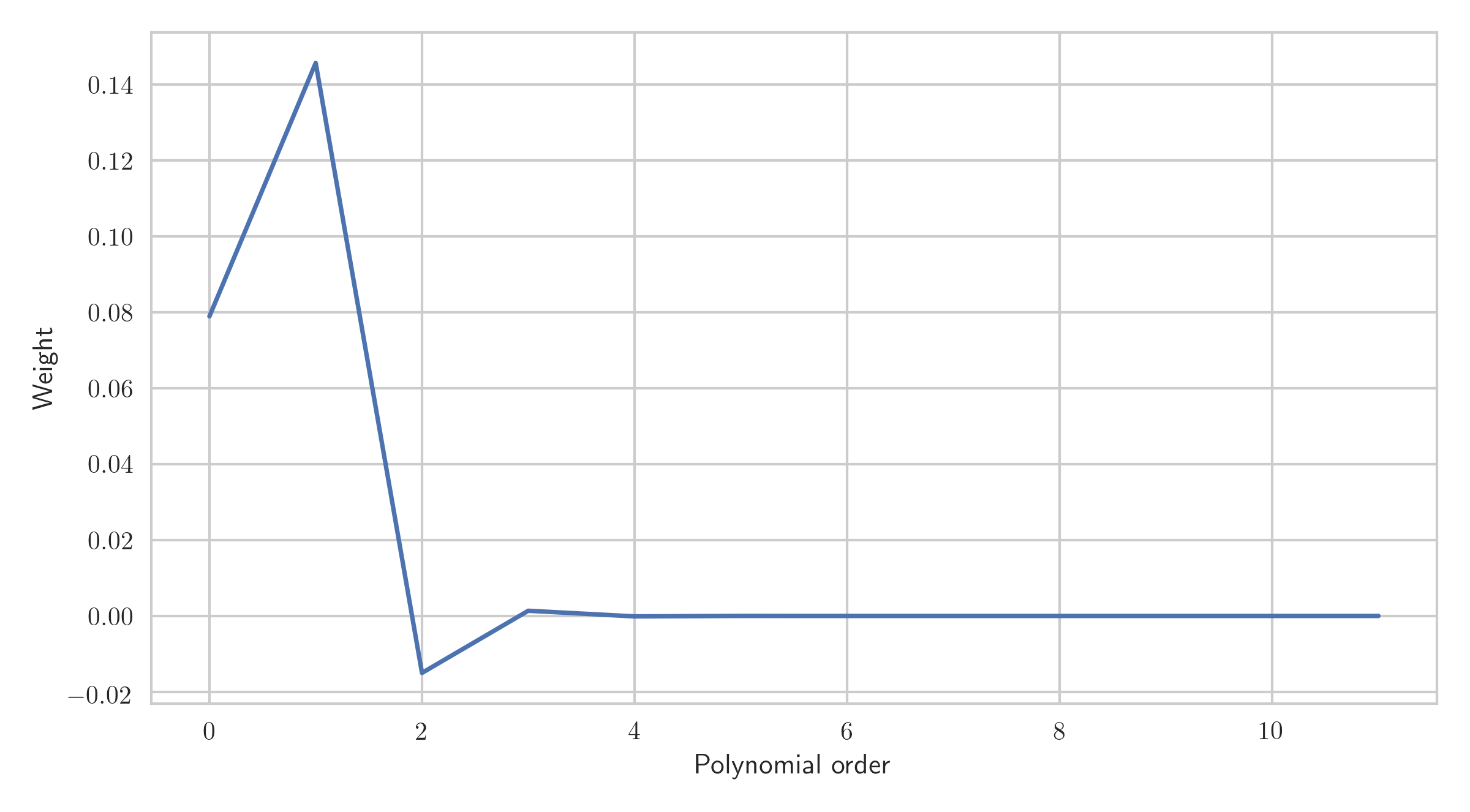}
    \end{center}
\caption{\label{fig:sintheta} The weights for an expansion of the $\sin^{2}(\theta(x))$,
  based on the angle to offset transform parameterized for typical interval ($v_{\rm int}$), and
  rms ($v_{\rm rms}$) velocities at the base of the Marcellus shale.}
\end{figure}

For the sake of computational convenience, we would look to reduce the
dimension of the parameter space by supposing that we can truncate the
polynomial expansion to some order.  Having decided where to truncate
the expansion, we will then require a method of interpretation given a
physical model. For our demonstration to the Marcellus shale, we will
analyze data with incident angles out to $35^\circ$. A suitable
physical model of the reflectivity on this domain is according to
\cite[]{Ruger:97}, in terms of the angle of incidence $\theta(x)$ and
azimuth $\phi$:
\begin{eqnarray}
  \label{eq:ruger1}
  r(x, \phi) &=& I +
                 ({\rm B}^{iso} + {\rm B}^{ani}\cos^{2}(\phi - \phi_{s}))\sin^{2}(\theta(x))\cr
                &=& I + ({\rm B}^{iso} + \frac{B^{ani}}{2})\sin^{2}(\theta(x)) + \
                     \frac{B^{ani}}{2} \cos(2\phi_{s})\cos(2\phi)\sin^{2}(\theta(x)) + \cr
                & &\hspace{3cm}     \frac{B^{ani}}{2} \sin(2\phi_{s})\sin(2\phi)\sin^{2}(\theta(x))\ .
\end{eqnarray}
The $\cos(2\phi),\ \sin(2\phi)$ dependence in the azimuth is orthogonal
to the harmonic functions with $l > 2$. By reciprocity, only even
coefficients of $\vert m \vert$ are non-zero. Now consider the angular
dependence by offset.  The simple angle-to-offset transform\cite[]{Castagna:93}:
\begin{eqnarray}
  \sin^{2}(\theta(x)) &=& \bigg(\frac{v_{\rm int}}{v_{\rm rms}}\bigg)^{2}\frac{x^{2}}{x^{2} + (v_{\rm rms}t_{0})^{2}}\ ,
\end{eqnarray}
(within approximately $30^{\circ}$), converges very rapidly in the basis of Legendre polynomials. 
The weights for the expansion:
\begin{eqnarray}
  \label{eq:weights}
  \sin^{2}(\theta(x)) &=& \sum_{i} w_{i}P_{i}(x), 
\end{eqnarray}
of even $x$, are shown in Fig.~\ref{fig:sintheta} for typical values of
rms- and interval velocity.  
With these constraints, it is computationally convenient to do our regression on
a re-summed expansion of the spherical harmonics in terms of the Legendre polynomials:
\begin{eqnarray}
  \label{eq:ruger2}  
  r(x,\phi) &=&  \sum_{i=0}\bigg(
                a_{i0}P_{i}(x) + 
                a_{i1}P_{i}(x)\sin(2\phi) + \ 
                a_{i2}P_{i}(x)\cos(2\phi)\bigg) \ .
\end{eqnarray}
Given the analysis shown in Fig.~\ref{fig:sintheta}, we will truncate the expansion
of the reflectivity in the basis of Legendre polynomials at 6-th order (inclusive),
using only the even polynomials. The attributes $a_{ij}$ are a set of coefficients
associated with each basis function, these basis functions are orthogonal. 

Combining Eqs.~\eqref{eq:ruger1}, \eqref{eq:weights} and \eqref{eq:ruger2} we have a table of
relationships:
\begin{eqnarray}
  \label{eq:ruger_interp}
a_{00} &=& I + \bigg({\rm B}^{iso} + \frac{B^{ani}}{2}\bigg)w_{0} \cr
a_{i0} &=& ({\rm B}^{iso} + \frac{B^{ani}}{2})w_{i} \cr
a_{i1} &=& \frac{B^{ani}}{2} \sin(2\phi_{s}) w_{i} \cr
a_{i2} &=& \frac{B^{ani}}{2} \cos(2\phi_{s}) w_{i} \ ,
\end{eqnarray}
the weights $w_{i}$ can be estimated by ray tracing, or using the
angle-to-offset transform.
From the relationships Eq.~\eqref{eq:ruger_interp} we can immediately
solve for azimuthal variables:
\begin{eqnarray}
  \label{eq:solns}
  \frac{\vert B^{ani} \vert}{2}\vert w_{i} \vert  &=& \sqrt{a_{i1}^{2} + a_{i2}^{2}} \cr
  2\phi_{s} &=& \arctan{\bigg(\frac{a_{i1}}{a_{i2}}\bigg)}\ ,
\end{eqnarray}
with the solution for isotropic intercept as:
\begin{eqnarray}
  I &=& a_{00} - a_{i0} \frac{w_{i}}{w_{0}} 
\end{eqnarray}
The solution for the isotropic gradient is somewhat ambiguous.  From
Eq.~\eqref{eq:ruger1}, we solve for the magnitude of the anisotropic
gradient, but can only estimate the orientation of the axis of
symmetry up to an overall phase of $90^\circ$. The isotropic gradient
cannot be unambiguously estimated without first knowing the sign of the
anisotropic gradient:
\begin{eqnarray}
  B_{\rm iso} w_{i} &=& a_{i0} \pm  \frac{\vert B^{ani} \vert}{2} w_{i}\cr
  {\rm or\ } && \cr
  B_{\rm iso} w_{i} &=& a_{i0} \pm  \bigg( a_{j1}\sin(2\phi_{s}) + a_{j2}\cos(2\phi_{s}) \bigg)\frac{w_{i}}{w_{j}} \ ,
\end{eqnarray}
While both of these equations are valid, in the presence of noise, the
distribution of $\vert B^{ani} \vert$ becomes skewed, and heavy
tailed. We expect that propagating this uncertainty will lead to a
skewed estimate of $B_{\rm iso}$. Finally, assuming stationarity in
$\phi_{s}$, we choose the solution for the isotropic gradient that is
consistent with the regional analysis of \cite[]{Schlanser:16}.



These relationships are valid for all $w_{i}\neq 0$, solutions for
$I$, $B_{iso}$, $B_{ani}$ and $\phi_{s}$ are determined by averaging
the appropriate normalized coefficients at over every order in the
expansion. To translate this result to our master formula
Eq.~\eqref{eq:posterior}, pack the basis
vectors:\\ $\{P_{0}(x_{i}),\ P_{0}(x_{i})\sin(2\phi_{i}),\ P_{0}(x_{i})\cos(2\phi_{i}),\ ...\}$
into the columns of matrix $A$ for each trace $i$ of the gather.
The rows of the solution $B$ contains the coefficients $a_{lm}$ at
each point in time. The coefficients $a_{lm}$ describe the reflectivity
surface at a particular time, a clustering algorithm is applied in
this space for a particular horizon. 

\section{Results}

\subsection{Synthetic Example}

We will first demonstrate the algorithm on a simple synthetic data
forward modeled using the Geokinetics' SynAVO package, with
modifications to simulate the azimuthal variation in reflectivity. The
SynAVO forward model used a 40Hz Ricker wavelet, with effects of
NMO-stretch included in the model.  Rock-properties for the reflectors
were take from the example Table 1 of \cite[]{Ruger:97}, with an
axis-of-symmetry at $30^\circ$. The sequence in each case is an
anisotropic material layered between two isotropic materials.  The
model simulates fluid filled cracks in a thin layer 75ft in thick,
surrounded by an isotropic material.
We simulate an OVT gather by forward modeling the seismic response
using a (10x5) grid of tiles, each separated by 350ft in each spatial
direction.

%
%
\begin{figure}
  \hspace{-1cm}
  \begin{tabular}{cc}
    \includegraphics[height=4.5cm]{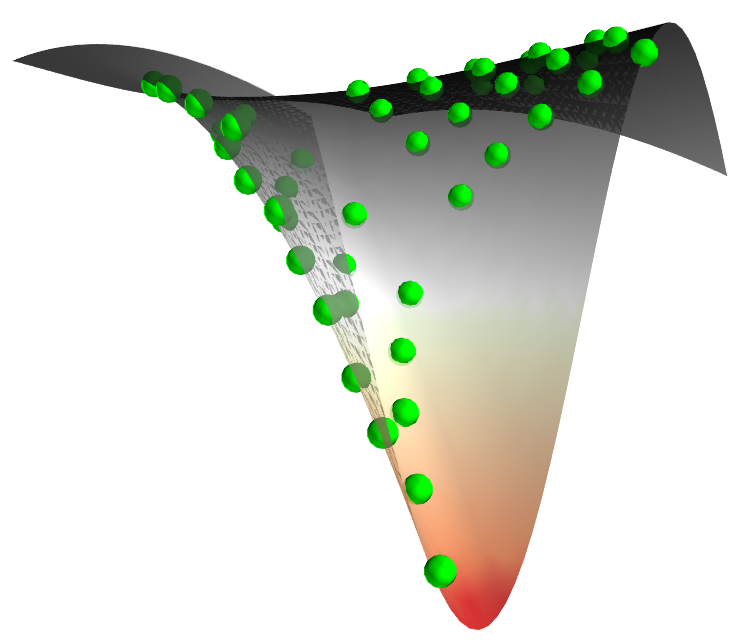} &
    \includegraphics[height=4.5cm]{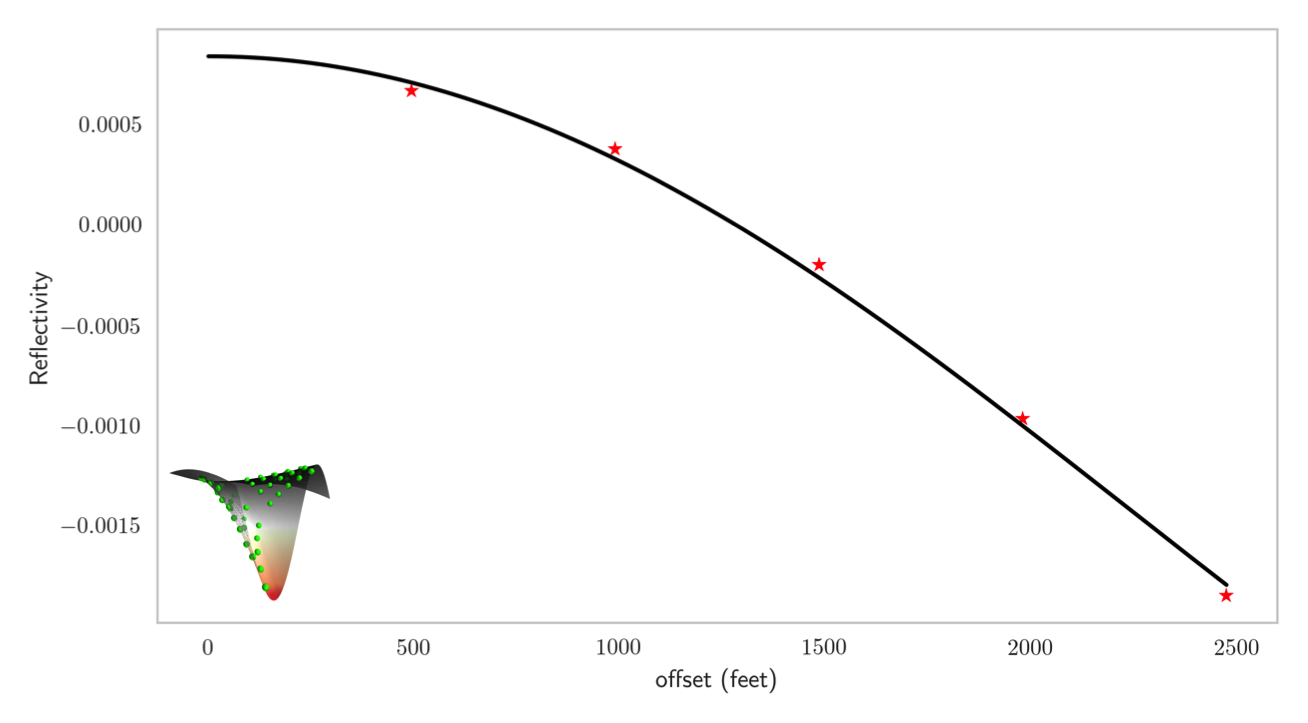} \cr
    \includegraphics[height=4.5cm]{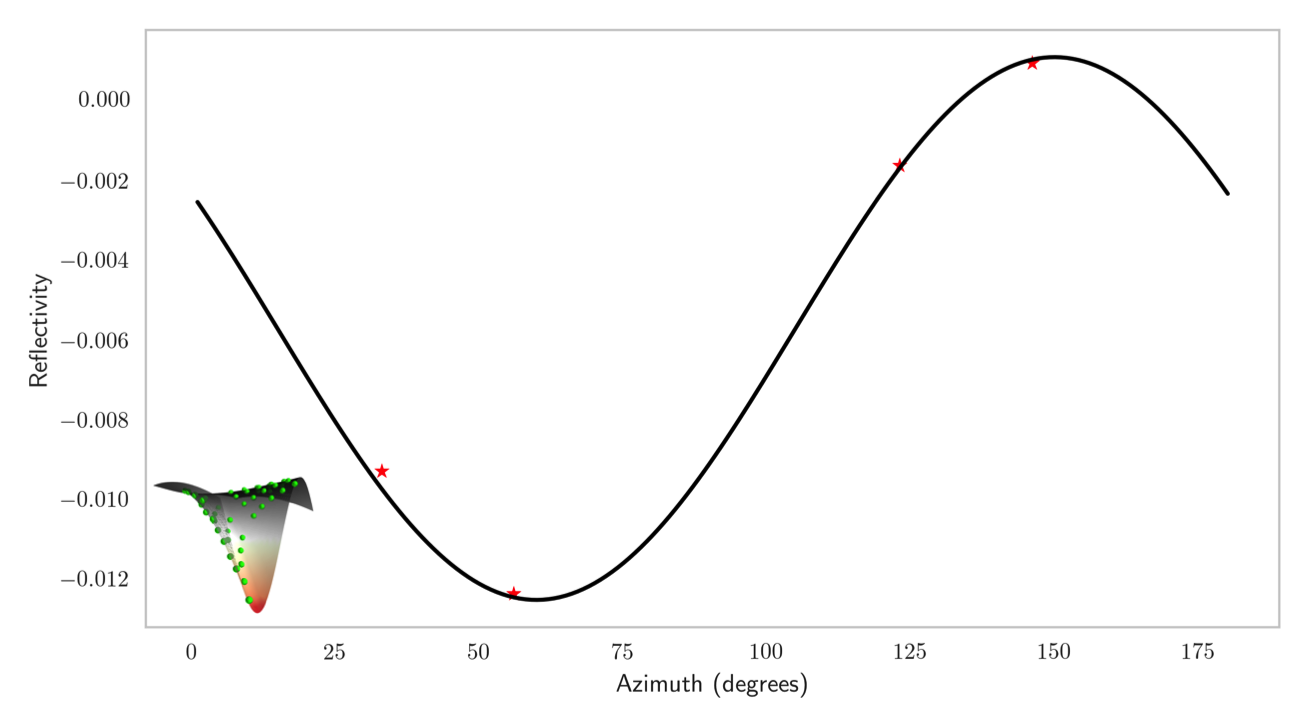} &    
    \includegraphics[height=4.5cm]{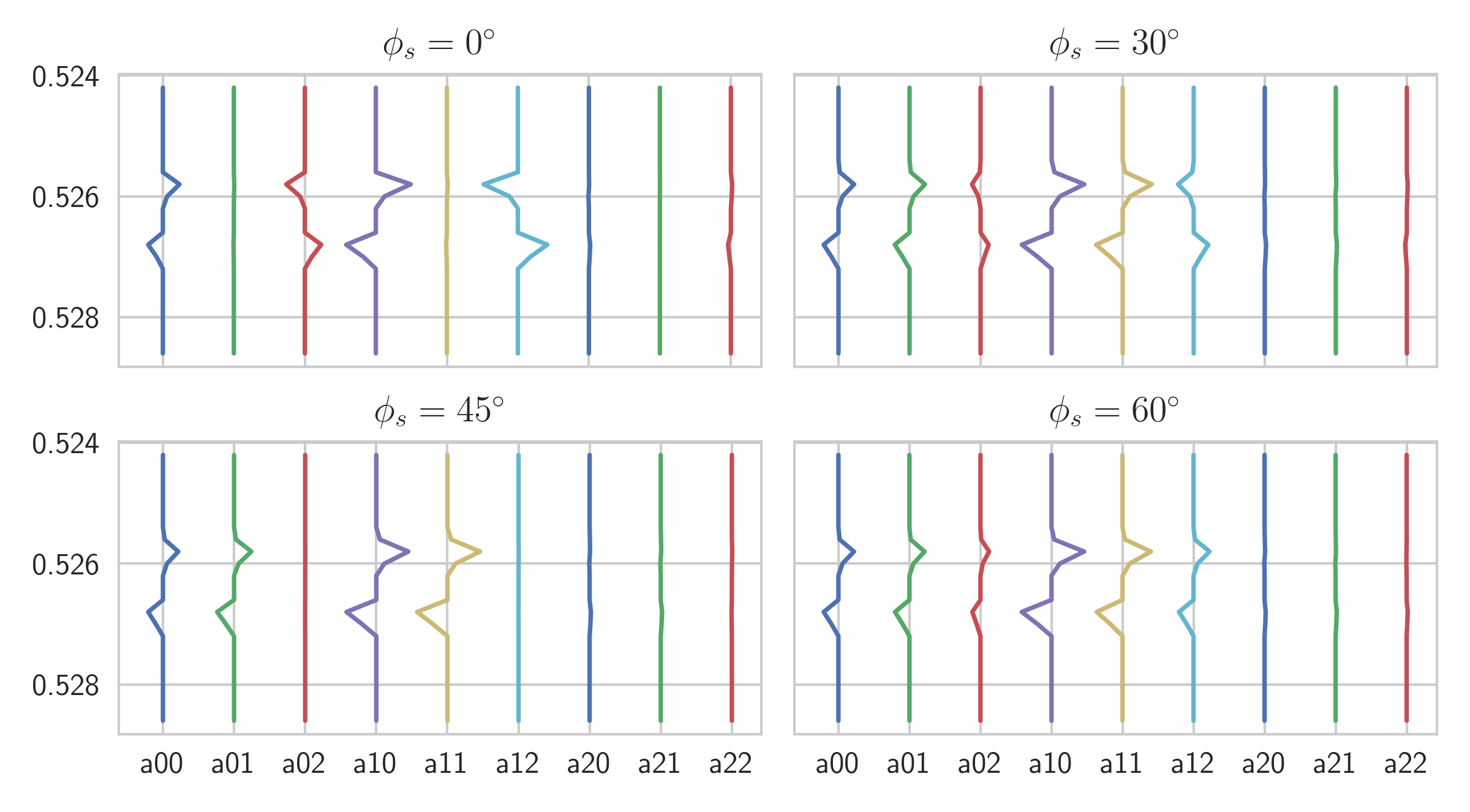} 
  \end{tabular}
  \caption{\label{fig:synth} (top left) Demonstrates the reflectivity
    surface inverted from the synthetic example at the base of the
    layer simulating fluid filled cracks. The vertical axis is
    amplitude, the horizontal axes are spatial. (green points) The
    amplitude of the seismic simulated by SynAVO at each offset and
    azimuth. (top right) At a fixed azimuth of $134^\circ$, (black
    line) the reconstructed seismic from the inversion as a function
    of offset. (red star) The synthetic data. (bottom left) Similarly,
    the reconstructed seismic and synthetic data at fixed offset of
    1250 feet. (bottom right) The spikes at each order of the
    expansion for the synthetic example, (clockwise) the
    axis-of-symmetry is rotated from $0^{\circ}$ to $30^{\circ}$,
    $60^{\circ}$ and $45^{\circ}$ respectively.  }
\end{figure}

The (green) points in Fig.~\ref{fig:synth} (top left) show the
amplitude of the synthetic gather as a function of offset and azimuth,
at the base of the simulated thin layer of wet cracks. The
optimization solves for attributes $a_{ij}$ using the methods
discussed above, where we have applied the grouped sparse constraint
only.  Combining the inverted attributes, with the polynomial basis
gives a a formula for the reflectivity surface Eq.~\eqref{eq:ruger2}
(a function of offset and azimuth). Convolving the reflectivity with
the wavelet provides a reconstruction of the seismic as a function of
offset and azimuth. This surface is overlayed on the image (top
left).
The dimensions of this image are amplitude (vertical axis) and spatial
(horizontal axes).
To further demonstrate the geometry of the problem, (top right)
and (bottom left) we show cut planes through the surface at a
particular offset and azimuth respectively. Here the (black line)
represented the reconstruction and (red star) shows the synthetic
data. The reconstruction is not exact because of the effect of
NMO-stretch, and because the optimization faces a trade-off between
minimizing both the residual variance and the complexity of the
solution. Where minimizing both absolutely is mutually exclusive.

The optimization framework we derive is very general, for a specific
application to real seismic data, we analyze the reflectivity
according to Ruger's model using the formula derived above.  In
Fig.~\ref{fig:synth} (bottom right) we show the inverted attributes
which appear as a sequence of sparse spikes. Here we resolve the top
and base of the reflector in the synthetic example.
The spikes have a finite bandwidth because the optimization cannot
identify the location of the reflector with resolution that exceeds
the Nyquist criterion. The amplitude of the reflectivity at a horizon
is found my integrating the spikes, a secant method, such
as~\cite[]{Glinsky:16} would be appropriate.
The first three attributes in each image are associated with the
$P_{0}(x)$, polynomial, the next three are associated with the
$P_{1}(x)$ polynomial, an so on. Our expectation for the relative
amplitude between these groups is shown in Fig.~\ref{fig:sintheta},
with the largest power associated with the $P_{1}(x)$ polynomial and a
very rapid convergence of the $\sin^{2}(\theta(x))$ dependence in the
basis of Legendre polynomials. From Eq.~\eqref{eq:ruger_interp}, the
amplitude of the $a_{00},\ a_{10},\ a_{20}$ attributes come from a
mixture of isotropic and anisotropic gradient, where the amplitude of
the $a_{i1},\ a_{i2}$ attributes originates solely from the presence of
the anisotropic gradient. Recall that $a_{i1},\ a_{i2}$ are
proportional to the sine and cosine of twice the axis-of-symmetry
($\phi_{s}$, relative to grid north). As the axis-of-symmetry rotates
in this example, we see the spectral content moving from the
$a_{i2}$ attributes at $\phi_{s}=0^\circ$, to the $a_{i1}$
attributes at $\phi_{s}=45^\circ$. Once the axis-of-symmetry reaches
$\phi_{s}=90^\circ$, the spectral content has rotated back to the
$a_{i2}$ attributes, which is indistinguishable from the
$\phi_{s}=0^\circ$.  This represents the so-called $90^\circ$
ambiguity in the two-term version of Ruger's model.

\subsection{Application to the Marcellus - Calibration}

In this section we will apply the sparse azimuthal inversion to field
data. The geological setting and aspects of the aquisition were
discussed previously.


To parameterize the model Eq.~\eqref{eq:posterior}, we need an
estimate of the seismic noise level and a value for the sparse
constraint, where we will use the same value for the group constraint
($\lambda_{g}$) for all time. Practically it is the product of the
prior rate parameter $\lambda_{g}$ and the noise-level $\sigma^{2}$
that will effect the result of the inversion, as such we only need a
approximate estimate of the noise level in order to have a good
initial guess of what the optimal value for $\lambda_{g}$ might
be. The median variance of a seismic trace computed from all gathers
within the dataset is used to set the background noise level.
A better estimate of the seismic noise level could also be estimated
jointly with the wavelet extraction using a Bayesian methodology such
as~\cite[]{Gunning:06}.

\begin{figure}
  \hspace{-1cm}
  \begin{tabular}{lr}
  \includegraphics[height=5.5cm]{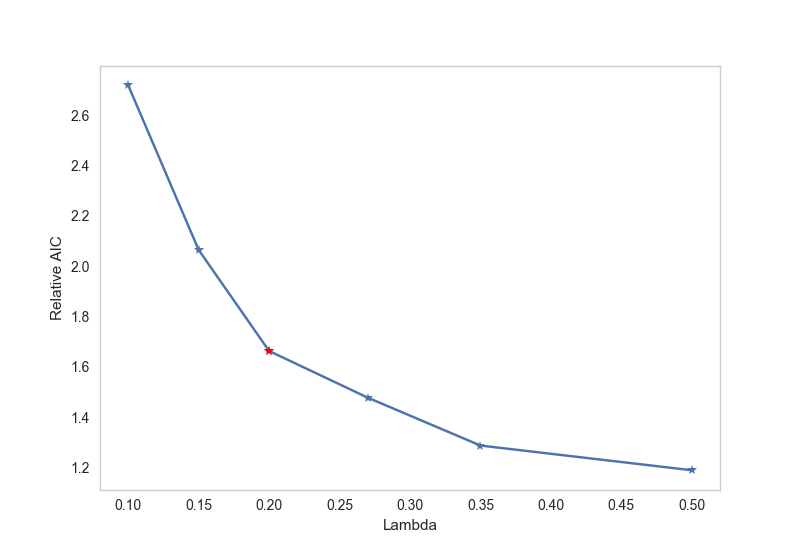} &    
  \includegraphics[height=5.5cm]{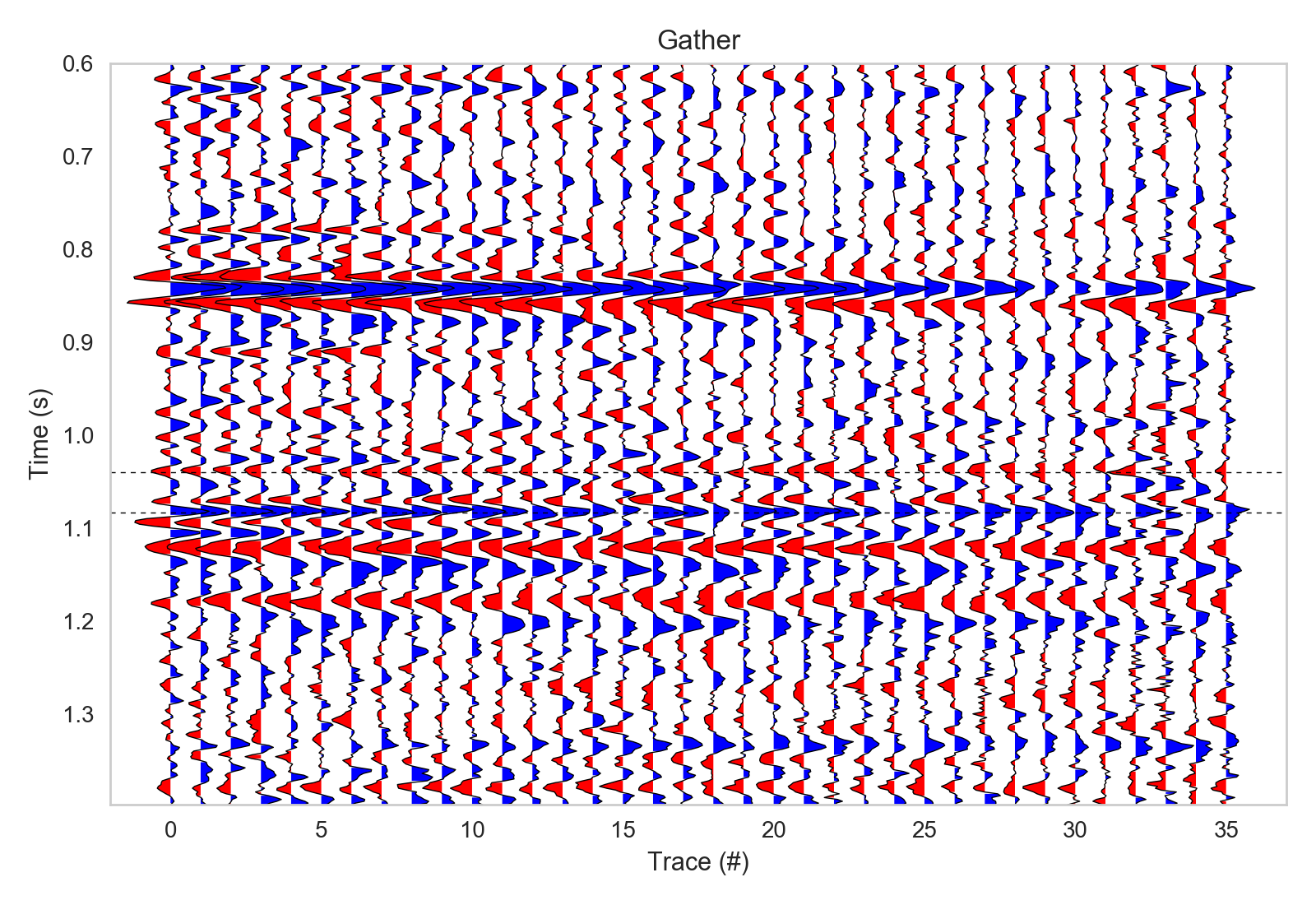} \cr
  \includegraphics[height=5.5cm]{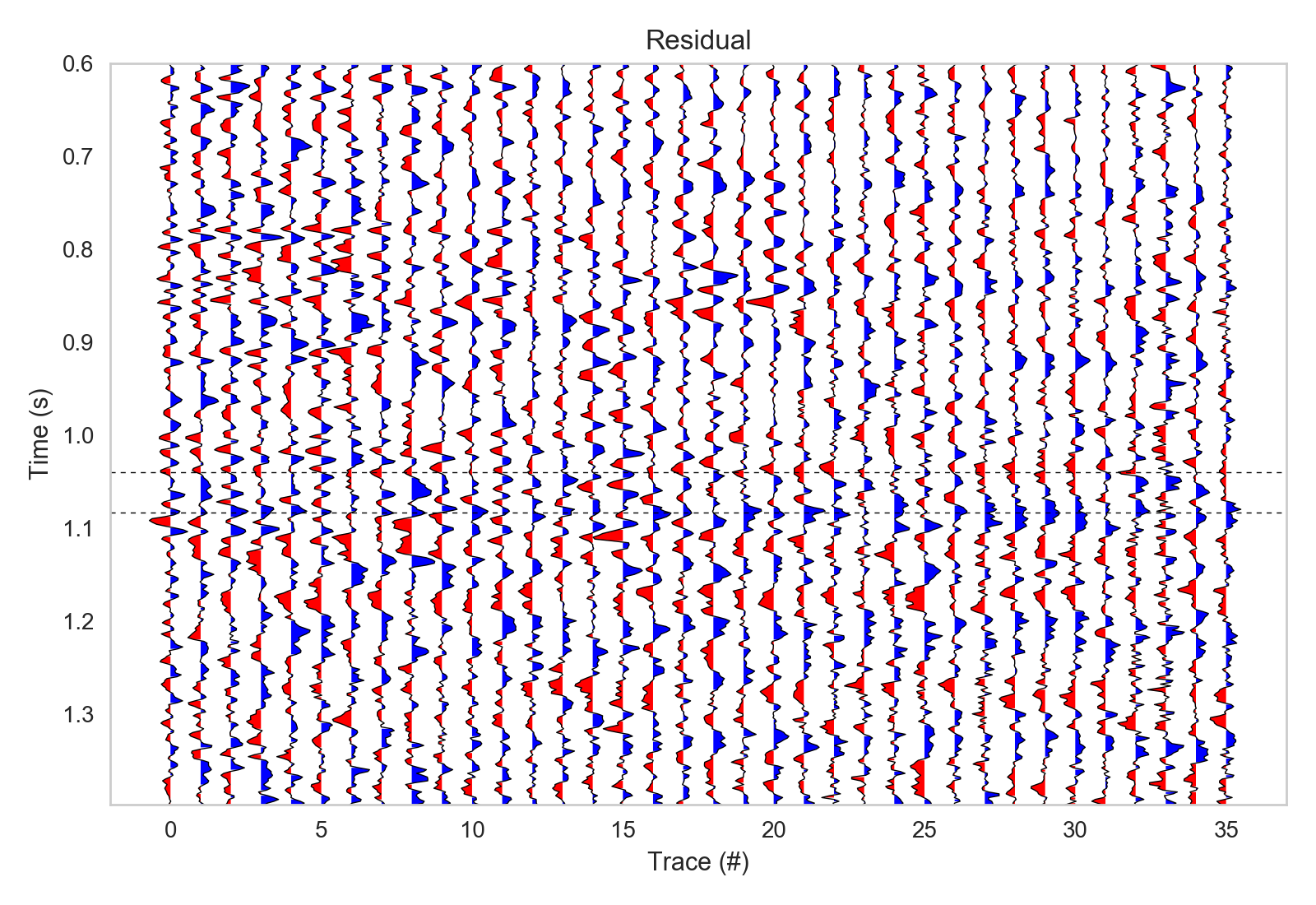} &
  \includegraphics[height=5.5cm]{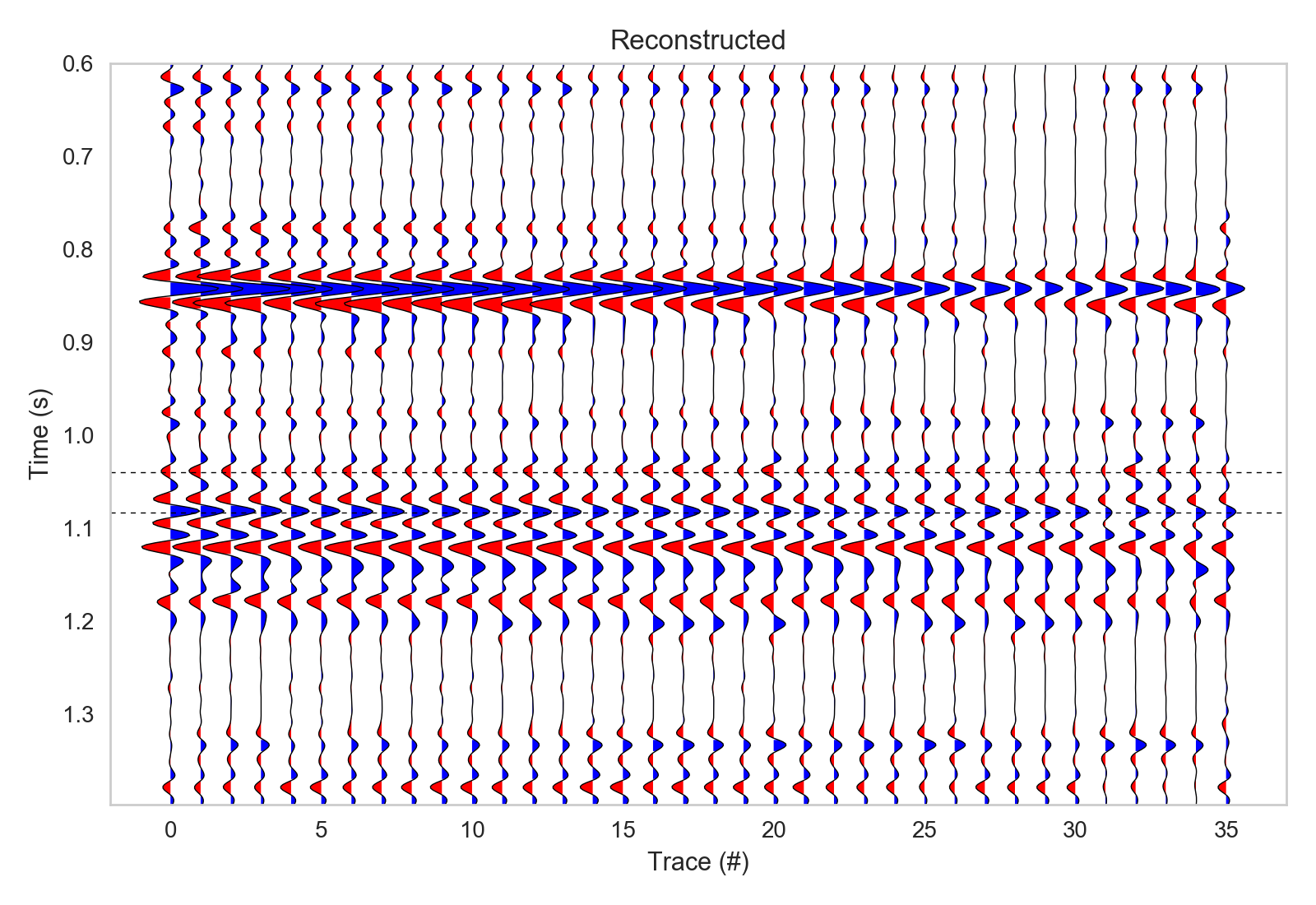} 
\end{tabular}
  \caption{\label{fig:gth6100} (top left) The relative
    Akaike information criteria for various values of the $\lambda_g$.
    (red star) The value of $\lambda_g$ used for the analysis throughout.
    (top right) An example OVT gather with traces ordered by offset and
    azimuth. (bottom right) The OVT gather reconstructed by convolving the
    extracted wavelet with the estimated reflectivity. (bottom left) The
    difference between the original and reconstructed gather.
  }
\end{figure}

Given this estimate of the noise level, we explore values
of $\lambda_{g}$ for a given inline, as shown in
Fig.~\ref{fig:gth6100} (top left). The goal is to find the least
complex model that can explain the variability in the data. In our
example there are two very prominent reflectors, which are the Tully and
Onondaga limestones.  To calibrate $\lambda_{g}$, we start with a
value that will resolve these features, and little else. Then allow
the complexity of the model to increase by reducing the value of
$\lambda_{g}$. The exponential of the relative Akaike information
criteria (AIC) for a series of values of $\lambda_{g}$ is shown in
Fig.~\ref{fig:gth6100}.  This curve reflects a trade-off between the
ability of the model to explain the variance in the data, and the
complexity of the model.  We choose the knee-point of this curve as
the ideal trade-off between these competing goals.

As a quality control, we will analyze a particular gather from the
dataset.  In Fig.~\ref{fig:gth6100} (top right) we show a the gather as
a function of time, where the traces are sorted by offset and azimuth,
note that the OVT data is not regularly sampled in this domain. The
inversion is then done with $\lambda_{g}=0.20$, the resulting spikes
are then re-convolved with the wavelet to give a reconstructed gather,
this is shown (bottom right). The residual difference between the
input and reconstructed gathers are shown (bottom left). From the
point of view of the model, the residual is random, band-limited
noise.

\begin{figure}
  \hspace{-1cm}
  \begin{tabular}{lr}  
  \includegraphics[height=5.5cm]{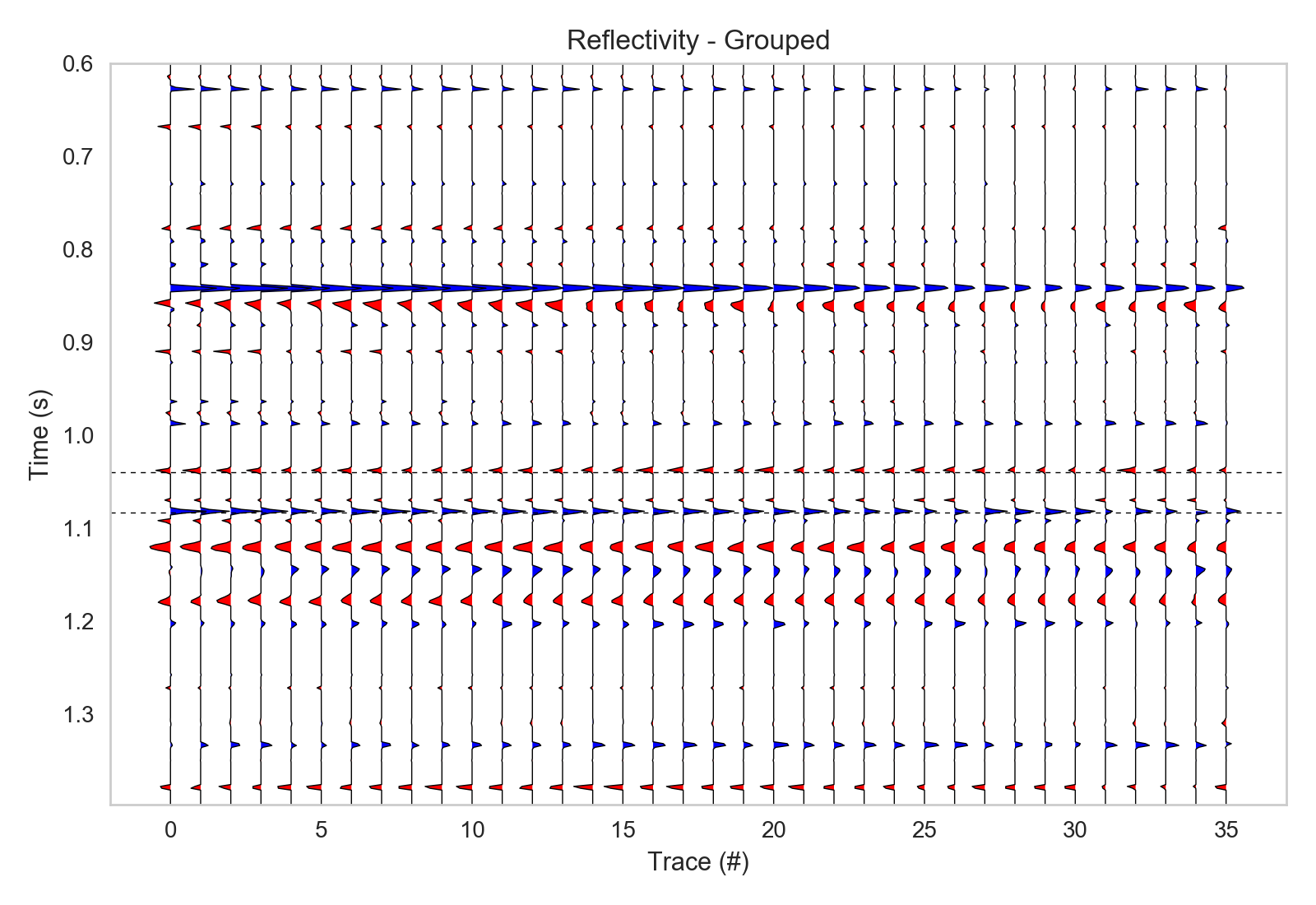} &
  \includegraphics[height=5.5cm]{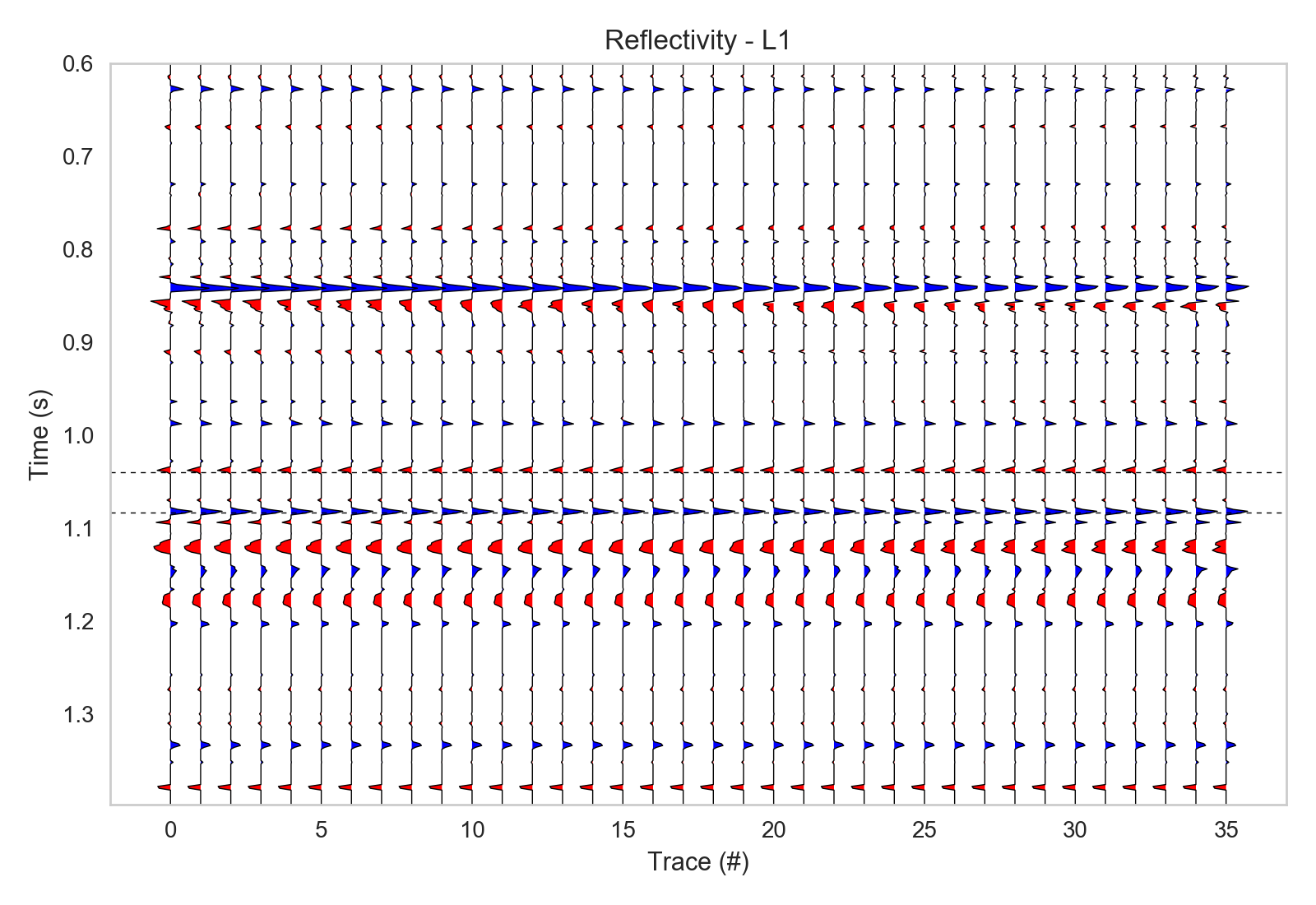} 
\end{tabular}
\caption{\label{fig:refl} (left) Reflectivity inverted from the
  gather shown in Fig.~\ref{fig:gth6100} using only the grouped
  constraint. (right) A hypothetical equivalent reflectivity
  inverted using a standard L1-constraint.}
\end{figure}

The inverted reflectivity form this analysis are shown in
Fig.~\ref{fig:refl} (left). For comparison, we plot the inverted
spikes for a model where we only use the global sparse constraint
(calibrated to produce an equivalent residual variance). The two
models produce a very similar picture of which reflectors are
important for describing variability in the data, however the grouped
constraint does not provide the same penalty on the model complexity
by offset and azimuth for a given reflector.

\begin{figure}
  \hspace{-1cm}
  \begin{tabular}{c}
    \includegraphics[height=8.5cm]{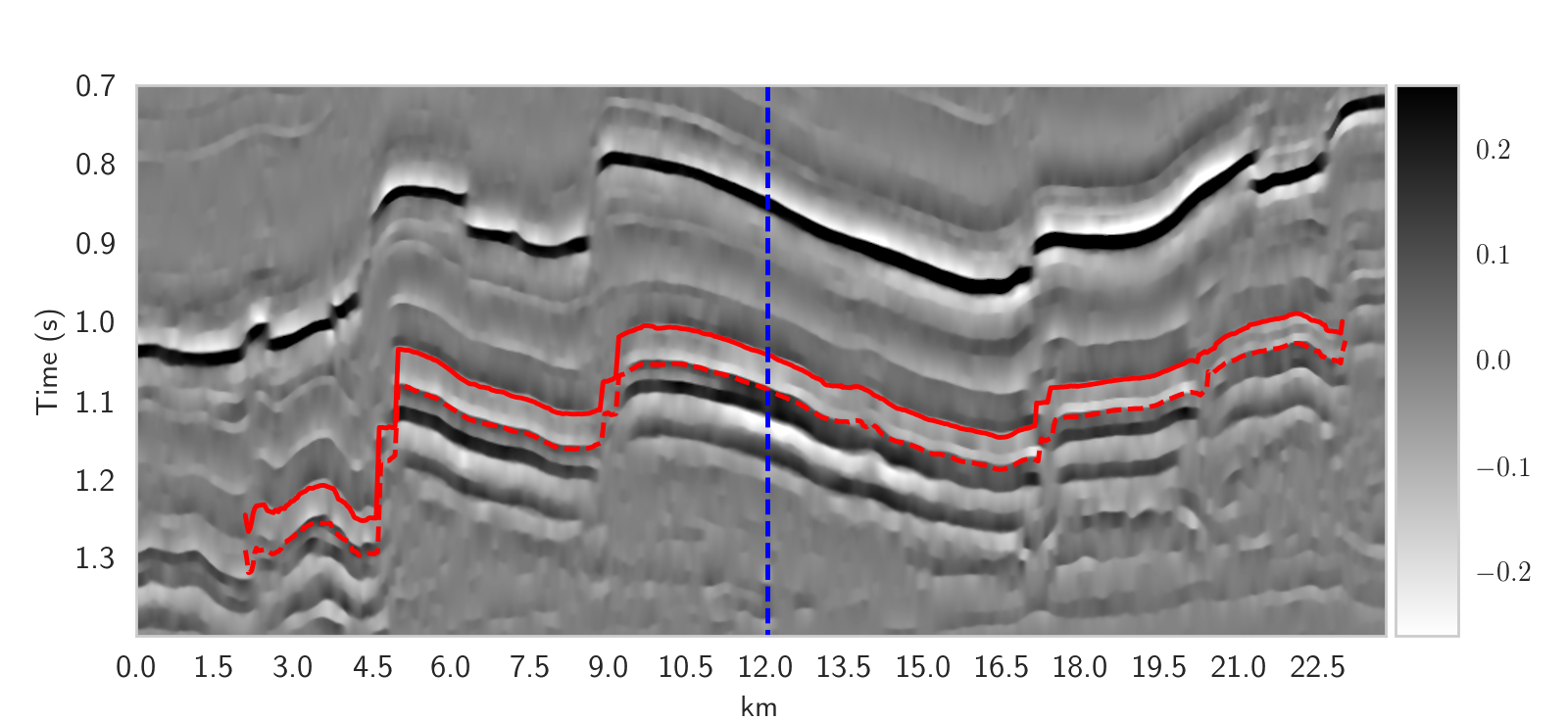}
    \end{tabular}
  \caption{\label{fig:target} (top left) A crude inversion for
    relative acoustic impedance on an inline.
    (dashed blue) Indicates the location of the gather used in
    calibration. (solid red) Line shows the horizon at the
    top of the Marcellus shale. (dashed line) Line shows the horizon
    at the top of the Onondaga limestone (base of the Marcellus)}
\end{figure}

Finally, the AIC provides a very general method for calibrating the model,
however there is some largess for the user to vary the model
complexity given an understanding of geology. In Fig.~\ref{fig:target},
we show a crude inversion of relative acoustic impedance derived
from the reflectivity Fig.~\ref{fig:refl} (left). Here we take
the cumulative sum of the inverted reflectivity at the nearest offset. 
We then apply a high-pass filter with a corner frequency of 6Hz.
The location of the example gather is shown with the vertical dashed line,
the picked horizons for the top and base of the Marcellus are shown as
(solid red) and (dashed red) lines respectively.
Interpretation of well-log data suggests the presence of the so-called
``Hot Marcellus'' in this area. The Hot Marcellus is a thin layer just
above the Onondaga limestone that is characterized by relatively high
total organic content compared to upper Marcellus. This is indicated
in the crude inversion by a dip in acoustic impedance, which we 
see evidence for in Fig.~\ref{fig:refl} (above the red dashed
line). We find that the statistically calibrated model $\lambda_g =
0.20$ is sufficiently complicated to reveal this feature.

\subsection{Application to the Marcellus - Unsupervised learning}
\label{ssec:learning}

Repeating the inversion for a set of gathers within a 3D seismic
survey, a hand picked horizon at the base of the Marcellus is used to
extract the coefficients of the expansion (Eq.~\eqref{eq:ruger2}) for
this reflector.  These coefficients represent a concise description of
the reflectivity surface at each point on the horizon. We can then
apply an unsupervised learning algorithm determine if these
coefficients are clustered in the parameter space. The EM-classifier
we use to do this requires us to specify the number of clusters
present. To determine this number we successively increase this number
from one, until the addition of a cluster does not appear to group the
data spatially. Using this method we arrived at three clusters, one of
which groups the data around faults where our picked horizon is less
reliable. We find that the performance of the classifier can be
improved by first normalizing the coefficients using the relevant
weights $w_{i}$ (computed from the angle-to-offset transform) to
adjust for differences in the angle of incidence across the horizon.

\begin{figure}
  \hspace{-1cm}
\begin{tabular}{lr}
  \includegraphics[height=4.5cm]{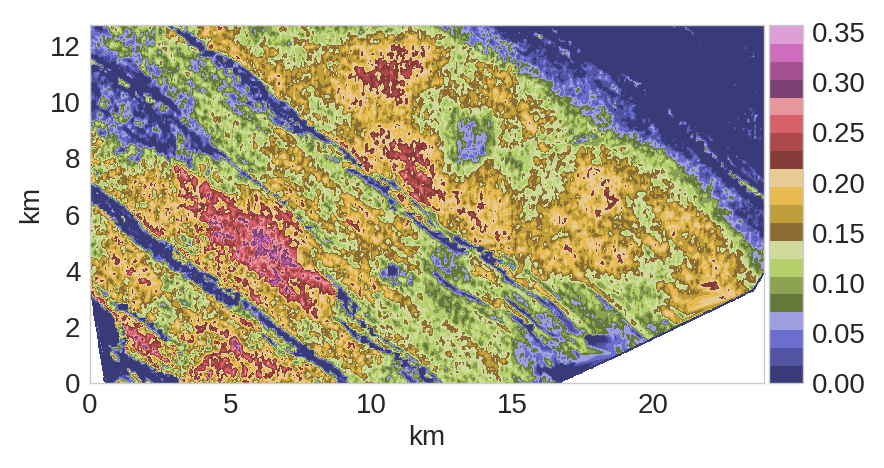} &
  \includegraphics[height=4.5cm]{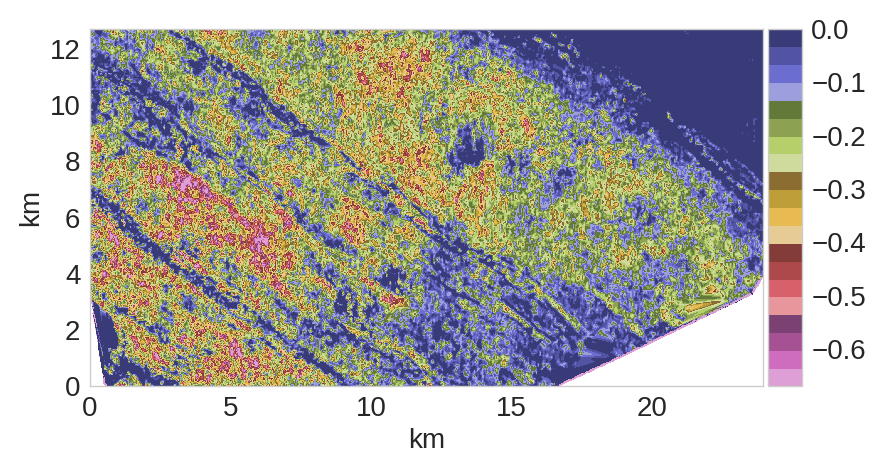} \cr
  \includegraphics[height=4.5cm]{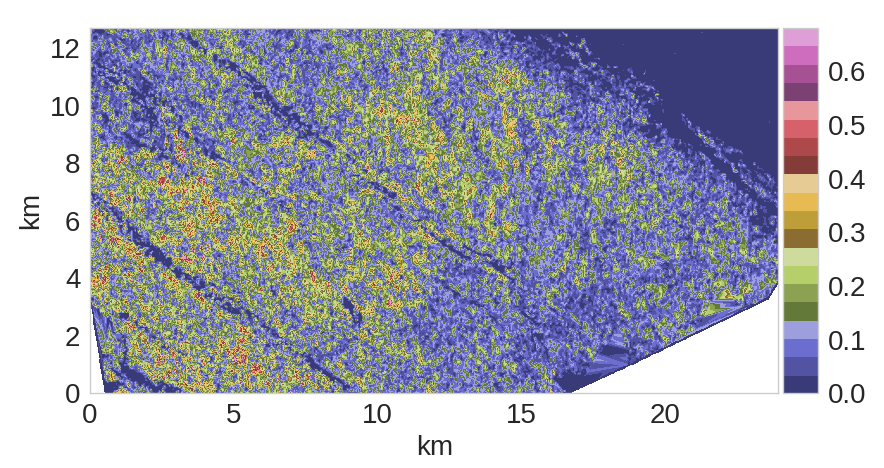} &
  \includegraphics[height=4.5cm]{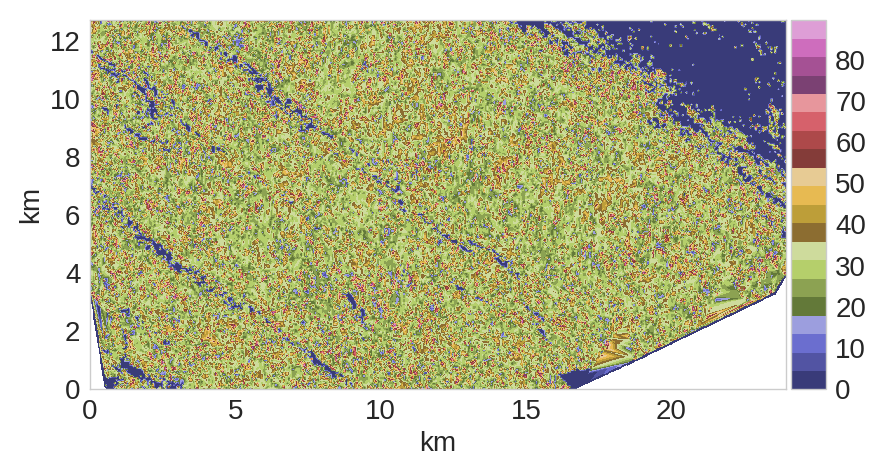} \cr
\end{tabular}
\caption{\label{fig:inv} Horizons at the base of the Marcellus showing
  (top left) the intercept, (top right) the isotropic gradient, (bottom left)
  the anisotropic gradient and (bottom right) the orientation of the
  axis of symmetry of the fractures.}
\end{figure}

\begin{figure}
 \hspace{-1cm}
  \begin{tabular}{lr}
  \includegraphics[height=4.5cm]{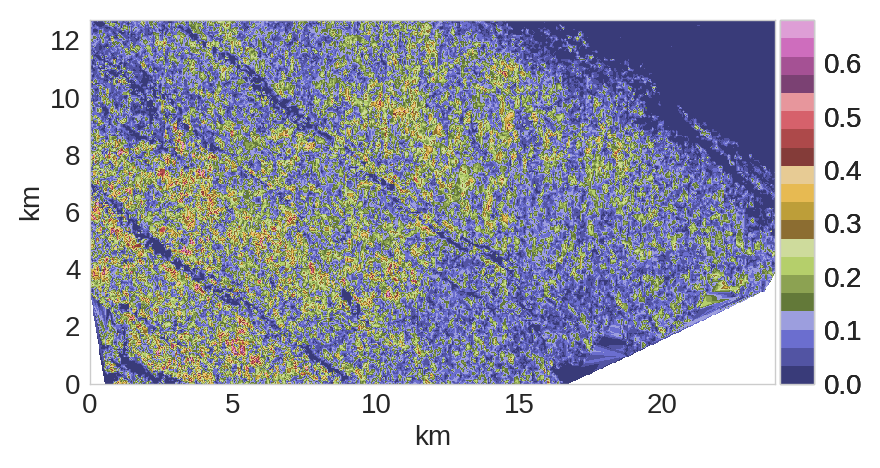} &
  \includegraphics[height=4.5cm]{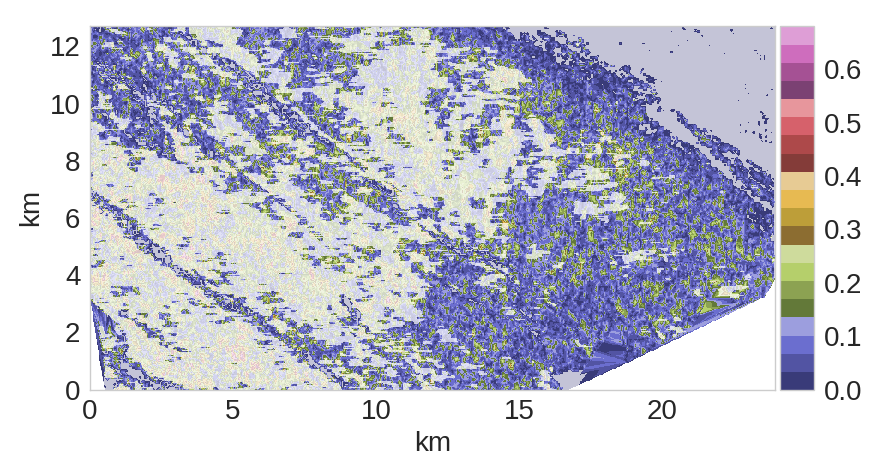} \cr
  \includegraphics[height=4.5cm]{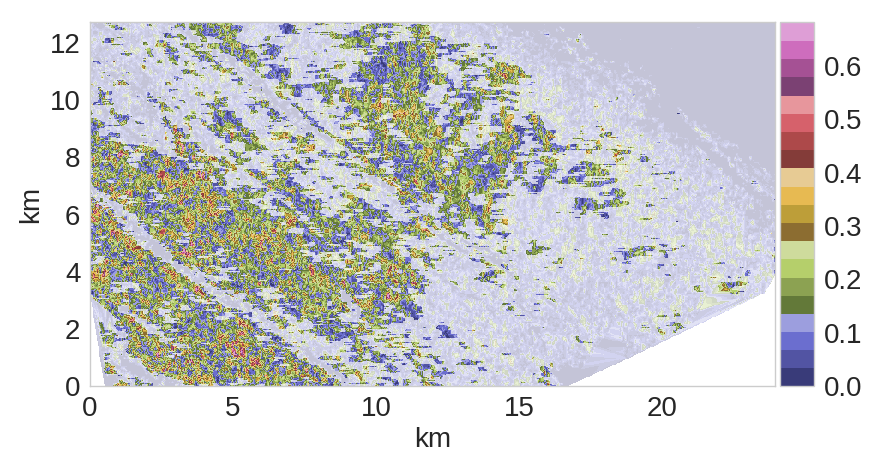} &
  \includegraphics[height=4.5cm]{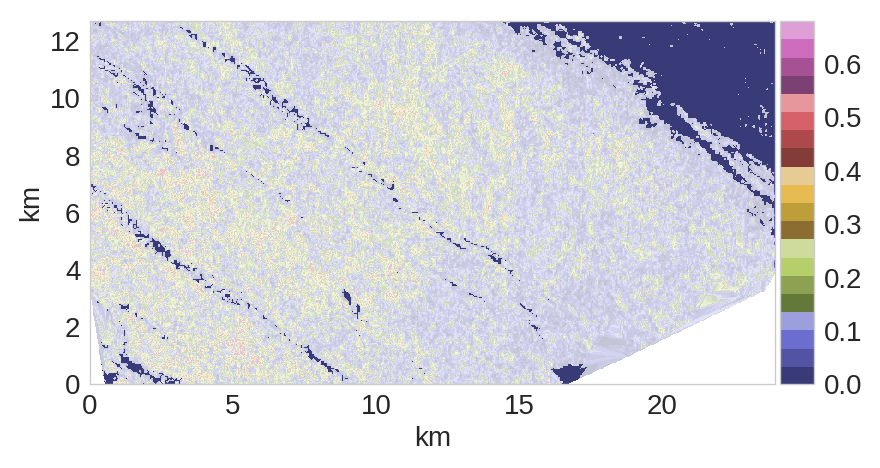} \cr
\end{tabular}
  \caption{\label{fig:cluster} For the horizon at the base of the
    Marcellus, showing (top left) The magnitude of the anisotropic
    gradient.  (otherwise) The magnitude of the anisotropic gradient
    for each cluster identified in the parameter space of the
    coefficients $a_{ij}$ defined in Eq.~\eqref{eq:ruger2}.  }
\end{figure}

Having identified spatial structure in the inverted coefficients, we
now have a problem of interpretation, for this we need to appeal to a
physical model.  The coefficients of Ruger's model are determined
using the method described above. The ambiguity in the sign is
resolved by ensuring that the isotropic gradient is consistent with
regional data derived from~\cite[]{Schlanser:16}.
The model parameters for base of Marcellus are shown in
Fig.~\ref{fig:inv}. The application of the clustering algorithm is
shown in Fig.~\ref{fig:cluster}, overlayed onto the magnitude of the
anisotropic gradient (that we may estimate unambiguously).

\begin{figure}
\begin{tabular}{cc}
  \includegraphics[height=5.5cm]{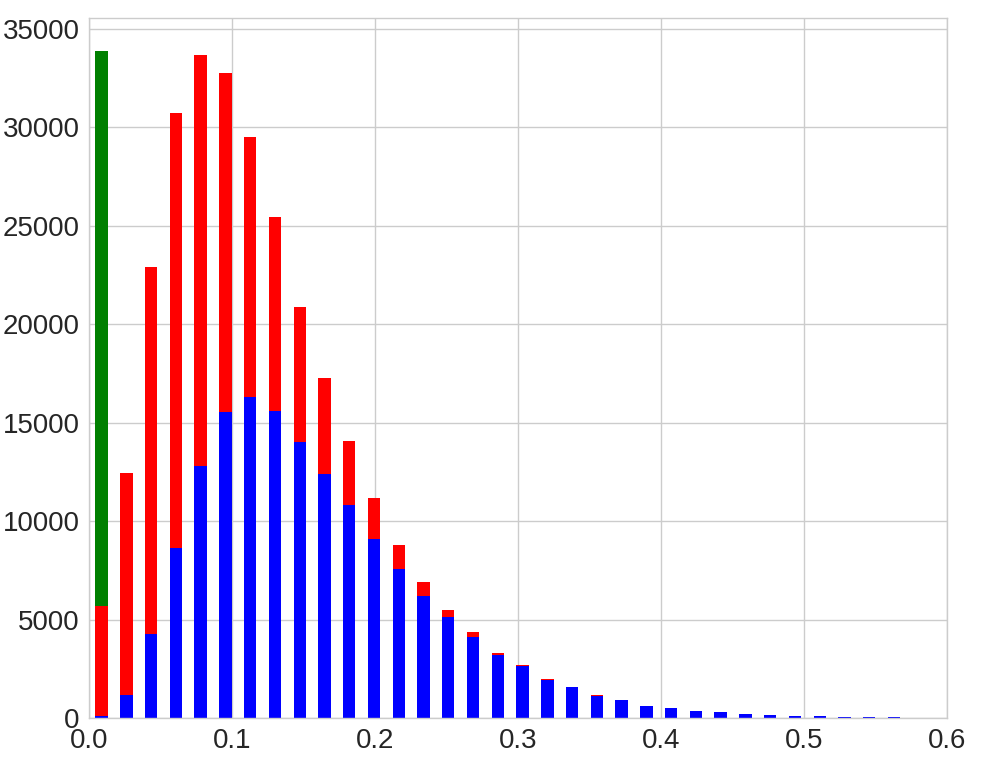} &
  \includegraphics[height=5.5cm]{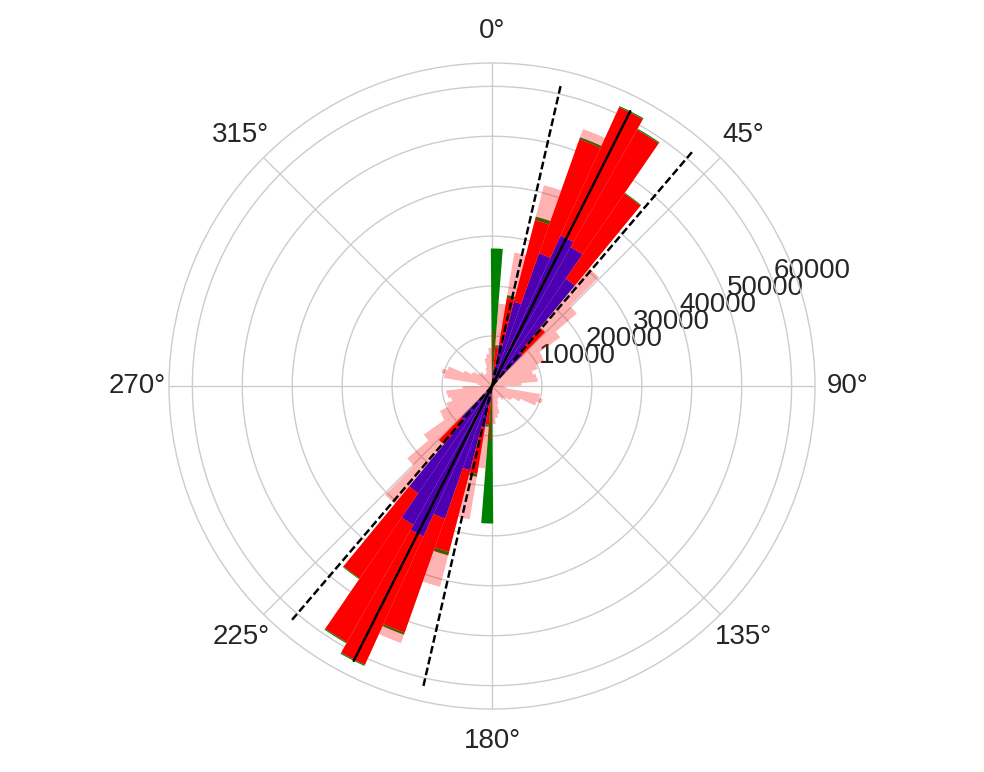} \cr
  \includegraphics[height=5.5cm]{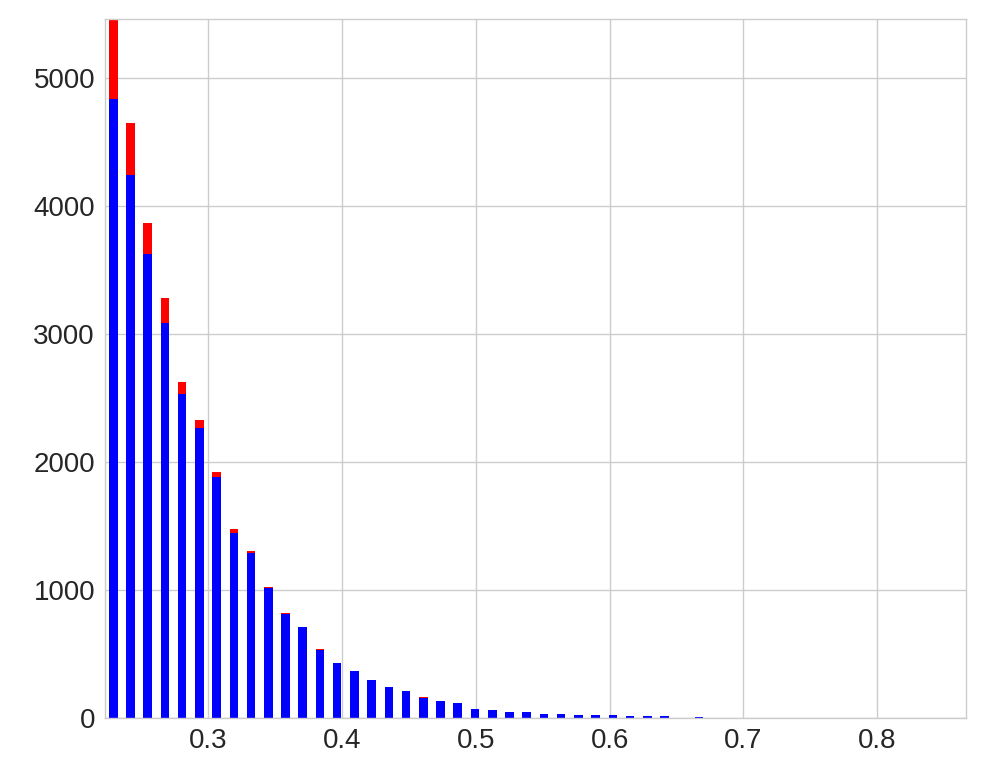} &
  \includegraphics[height=5.5cm]{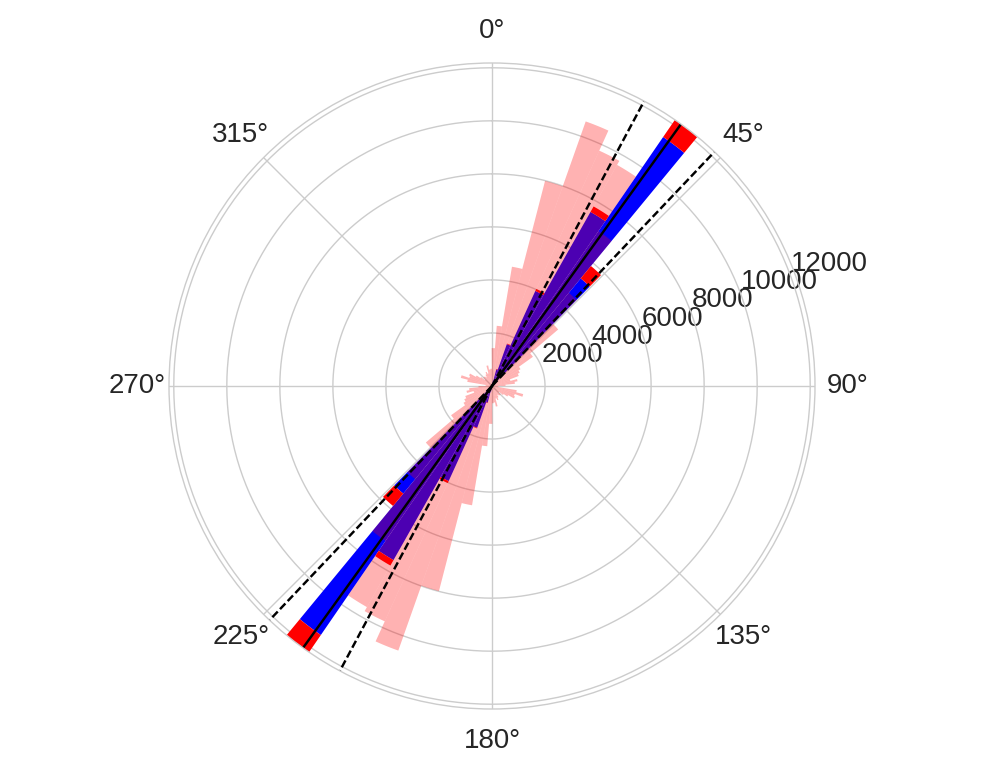} \cr
\end{tabular}
\caption{\label{fig:distro} For the horizon at the base of the
  Marcellus. (top left) The distribution of the magnitude of
  anisotropic gradient. (bottom left) The same
  distribution magnifying the top decile sorted by the magnitude of
  anisotropic gradient. (top right) The distribution of estimates of
  the axis-of-symmetry, overlayed (light red) the distribution of the
  axis-of-symmetry estimated using the HTI-correction. (bottom right)
  The distribution of the axis-of-symmetry for the top decile of
  points sorted by the magnitude of anisotropic gradient.}
\end{figure}

In Fig.~\ref{fig:distro} (right top and bottom) we show a histograms
of the phase and magnitude of the anisotropic gradient for each class,
and the distribution for the top decile of anisotropic gradient
respectively. We see that the tail of the distribution is dominated by
a single class. Interpreting one class as capturing areas around
faults and at the periphery of the survey, the other two are
characterized evidence of fracturing. Finally in Fig.~\ref{fig:distro}
(left top and bottom) we show the distribution of $\phi_{s}$ (up to
$90^{\circ}$), compared to the direction of the fast HTI velocity
(overlaid), this data is taken from the HTI time migration.
We see good agreement
between the two analyses.

\begin{figure}
  \begin{center}
    \includegraphics[height=6.5cm]{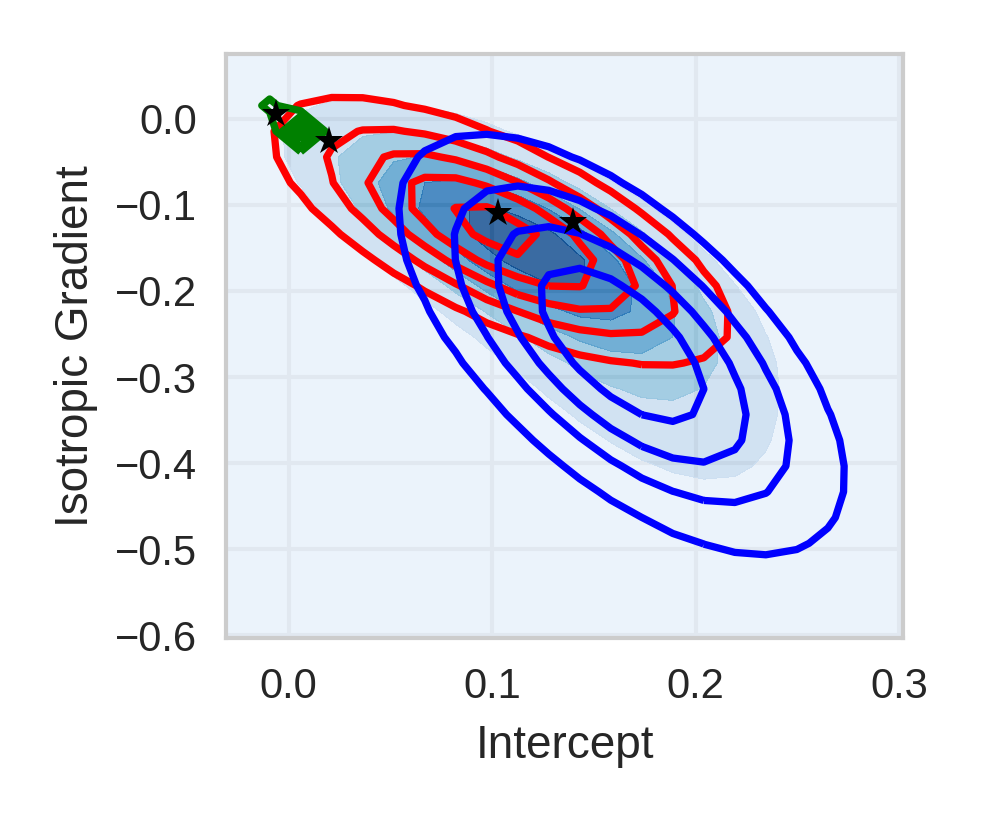}
    \end{center}
  \caption{\label{fig:cross} The distribution of the intercept and isotropic gradient
    values for each class. (blue) The distribution of
    intercept and isotropic gradient for the fractured class, (red) similarly for the
    unfractured class. The (green) class absorbs the zero and outlier data.
    (black star) Marks the modeling intercept and isotropic gradient using elastic
    properties derived for the gray shale, dark gray shale, black shale and
    calcareous from \cite[]{Schlanser:16}. The interface is assumed to be the carbonate,
    with properties derived from the same paper. 
  }
\end{figure}

Finally, in Fig.~\ref{fig:cross} we cross-plot the distribution of
intercept and gradient terms for each class. The (black star) marks
forward modeled values of intercept and gradient derived from the
P-wave velocity and density values for the gray-, dark gray-, black-
and calcareous-shales interfacing with a carbonate. The values used
were taken from \cite[]{Schlanser:16} for the Northern Pennsylvania
region (Zone A). This data was derived from multiple well-logs in the
zone of interest. Since S-wave velocity is not supplied in
\cite[]{Schlanser:16}, we used the Castagna-Greenberg relations to
model this for the shales and carbonates
respectively~\cite[]{Mavko:09}.
As expected, that the range of isotropic gradient is greater for
the class we associate with the fracturing. One likely explanation
for this is that the variance in the estimate of the isotropic
gradient is additive with the variance in the estimate of the
anisotropic gradient, contaminating the estimate of isotropic
gradient with additional noise. For the unfractured class, the center
of the distribution was consistent with the intercept and gradient
derived from the high TOC dark gray- and black-shales, layered
on top of the carbonate. This is consistent with our observation
of the ``Hot Marcellus'', a high TOC shale, that we saw evidence
for in Fig.~\ref{fig:gth6100}.

\section{Conclusions}

In this paper we derived a method for estimating independent
attributes describing reflectivity by offset and azimuth. The
inversion was subject to very general prior assumptions about sparsity
of the layered earth and smoothness of the reflectivity
surface. Otherwise we did not presume that any particular physical
model should apply, but let the data drive the inversion. Given an
inverted set of independent attributes we then applied a clustering
algorithm to classify the parameter space for a horizon at the base of
the Marcellus shale. To improve the robustness of the statistical
analysis, our method was derived to take common-offset or offset
vector tiles as direct input, without the need for interpolation or
fold-compensation that is required for processing sectored azimuthal
gathers. We argue that the assumptions of sparsity and smoothness are
guaranteed to be AVO compliant, and so reduce potential sources of
bias from the processing.

We then presented an analysis of the Marcellus shale as an example
application of our workflow. The target of the analysis was a horizon
at the base of the Marcellus shale. We estimated AVO attributes 
encoding our knowledge of the reflectivity without reference to a
specific physical model. We then applied an unsupervised learning
algorithm (EM algorithm) in this space of attributes. Interpreting the
result in the context of Ruger's model, we computed the magnitude of
the anisotropic gradient and compared the populations of the learned
classes. Sorting the observed magnitude of anisotropic gradient, we
saw that the top decile of values could be attributed to one class in
particular. 
%
We also compared the estimated axis-of-symmetry $\phi_{s}$ using
Ruger's model to the fast velocity direction estimated from the
HTI-migration. We can only resolve the orientation of $\phi_{s}$ to
within $90^\circ$, however we found that it was consistent with the
HTI-velocity analysis.

We then compared the estimates of the intercept and isotropic gradient
to values forward modeled from well-logs in the region. Regional models
where derived for four types of shale layered on top of a limestone.
Models of the two high TOC dark gray- and black-shales lay at the
center of cross-plots of inverted intercept and isotropic gradient
for the Marcellus shale dataset we analyzed. The distribution of
isotropic gradient was broader for the class associated with the large
values of anisotropy. We note that since estimates of intercept, isotropic- and
anisotropic-gradients are not made independently, uncertainty in
the estimate of anisotropic gradient propagates through to the
estimates of intercept and isotropic gradient.


\append{Treatment of non-orthogonal bases functions}
\label{app:qr}

The master equation Eq.~\ref{eq:master} assumes orthogonality of the
basis vectors packed into the columns of the matrix
$\gmat{A}$. Irregular sampling of the data spatially, or a preference
for a solution in a non-orthogonal basis breaks this constraint. In
this case we need instead solve a preconditioned version of the
problem.

The QR decomposition of the matrix $\gmat{A}$ can be written,
\[
\gmat{A} =
\left[
\begin{array}{cc}
\gmat{Q}_{11} & \gmat{Q}_{22} \\
\end{array}
\right]
\left[
\begin{array}{cc}
\gmat{R}_{11} & \gmat{R}_{12} \\
\gmat{0} & \gmat{R}_{22}
\end{array}
\right]
\]
where $\gmat{Q}_{11}$ is (N$\times$r) and $\gmat{R}_{11}$ is (r$\times$r). Provided the
matrix of basis functions $\gmat{A}_{\bot}$ is full rank, the above expression is
true if we define $\gmat{Q}_{11}$ and $\gmat{R}_{11}$ according to its QR factorization:
\[
\gmat{A}_{\bot} = \gmat{Q}_{11}
\left[
\begin{array}{c}
\gmat{R}_{11} \\
\gmat{0} 
\end{array}
\right]
\]
The residual from Eq.~\eqref{eq:master} can be expressed as, 
\[
\gmat{Y} - \gmat{W}
\left[
\begin{array}{cc}
\gmat{B}_{\bot} & \gmat{0} \\
\end{array}
\right]
\left[
\begin{array}{cc}
\gmat{R}^{T}_{11} & \gmat{0} \\
\gmat{R}^{T}_{12} & \gmat{R}^{T}_{22}
\end{array}
\right]
\left[
\begin{array}{c}
\gmat{Q}^{T}_{11} \\
\gmat{Q}^{T}_{22} 
\end{array}
\right]
\]
By post multiplying by $[\gmat{Q}_{11}, \gmat{Q}_{22}]$, 
\[
\left[
\begin{array}{cc}  
  \gmat{Y}\gmat{Q}_{11} & \gmat{Y}\gmat{Q}_{22}
\end{array}
\right]
- \gmat{W}
\left[
\begin{array}{cc}
\gmat{B}_{\bot}\gmat{R}^{T}_{11} & \gmat{0} \\
\end{array}
\right]
\]
Since the matrix $\gmat{Q}^{T}\gmat{Q} = \mathcal{I}_{N}$, we can solve the
preconditioned problem by regressing $\gmat{B}_{\bot}\gmat{R}^{T}_{11}$ on
$\gmat{Y}\gmat{Q}_{11}$
using Eq.~\eqref{eq:posterior} and then estimate the $\gmat{B}_{\bot}$ provided
the inverse of the (r$\times$r) matrix $\gmat{R}_{11}$ exists. This
requirement means that we cannot allow any columns of $\gmat{A}$ to be exactly
zero, and that each column cannot be linear combination of the
others. The problem will be ill-conditioned numerically if the columns
of $\gmat{A}$ are nearly co-linear, this would be the case for example, where
we used a $\sin^2$, $\tan^{2}$ basis for small angles.

\append{Generalized mixed constraint}
\label{app:mixed}

In this appendix we combine the concepts of Daubechies~\cite[]{Daub:04},
with the group lasso model of Chen\cite[]{Chen:12} and Friedman~\cite[]{Friedman:10}, 
to create a generalized algorithm for optimizing a function with mixed sparsity constraints.
Required normalization:
\begin{eqnarray}
  \Vert \gmat{W}^{T} \gmat{W} \Vert = 1\cr
  \gmat{A}^{T}\gmat{A} = \mathcal{I}_{N}\ ,
\end{eqnarray}
we will also suppose that we have re-scaled Y and W such that $\Sigma$ and $Q$ are the identity.
The cost function to be optimized is:
\begin{eqnarray}
  \mathcal{C}_{0}(\gmat{B}) &=& \Vert \gmat{Y} - \gmat{W}\gmat{B}\gmat{A}^{T}\Vert^{2}_{2}\cr
                     &=& \Vert \gmat{Y}\gmat{A} - \gmat{W}\gmat{B} \Vert^{2}_{2}\ ,
\end{eqnarray}
subject to the penalties:
\begin{eqnarray}
\mathcal{P}_{0}(\gmat{B}) &=& \lambda \Vert \Vect{\gmat{B}} \Vert_{1} \cr
\mathcal{P}_{1}(\gmat{B}) &=& \sum_{l=1}^{T} \lambda_{l} \Vert \gmat{B}^{l} \Vert_{2} \ .
\end{eqnarray}

A sub-gradient strategy for optimizing this cost function subject to
this mixed constraint is reviewed in \cite[]{Friedman:10}. We also have
an special case implementation of a very similar problem discussed
in\cite[]{Sassen:15}, this method takes advantage of the surrogate
functional procedure discussed by Daubechies\cite[]{Daub:04}. In our
experience this approach has very good convergence properties for the
application to AVO analysis of geophysical data. 

Following Daubechies\cite[]{Daub:04}, change the optimization problem
by adding a surrogate functional:
\begin{eqnarray}
  \mathcal{C}_{1}(\gmat{B}) &=& \Vert \gmat{B} - \tilde{\gmat{B}} \Vert^{2}_{2} -
  \Vert \gmat{W}\gmat{B} - \gmat{W}\tilde{\gmat{B}}\Vert^{2}_{2}\ ,
\end{eqnarray}
where we successively update our estimate of $\tilde{B}$ as the algorithm
iterates. The cost function to be optimized becomes:
\begin{eqnarray}
  \mathcal{C}(\gmat{B}) &=& \mathcal{C}_{0}(\gmat{B}) + \mathcal{C}_{1}(\gmat{B}) +
                           \mathcal{P}_{0}(\gmat{B}) + \mathcal{P}_{1}(\gmat{B})\ .
\end{eqnarray}
Evaluating the matrix derivatives of this function is straight forward:
\begin{eqnarray}
  \mathcal{C}_{0}(\gmat{B}) + \mathcal{C}_{1}(\gmat{B}) &=& \Vert \gmat{B} \Vert^{2}_{2} - \
  2\Tr{(\gmat{W}^{T}\gmat{Y}\gmat{A} - \gmat{W}^{T}\gmat{W} \tilde{\gmat{B}} + \tilde{\gmat{B}})^{T}\gmat{B}} + \
  ({\rm consts\ in\ \gmat{B}}) \cr
  {\rm with\ } \gmat{\Gamma} &=& (\gmat{W}^{T}\gmat{Y}\gmat{A} - \gmat{W}^{T}\gmat{W} \tilde{\gmat{B}} + \tilde{\gmat{B}}) \cr
  \frac{\delta}{\delta \gmat{B}} (\mathcal{C}_{0}(\gmat{B}) + \mathcal{C}_{1}(\gmat{B})) &=& 2 \gmat{B} - 2 \gmat{\Gamma} \cr
  \frac{\delta}{\delta \gmat{B}^{l}} \mathcal{P}_{0}(\gmat{B}) &=& \lambda \Sign{\gmat{B}^{l}} \cr
  \frac{\delta}{\delta \gmat{B}^{l}} \mathcal{P}_{1}(\gmat{B}) &=& \lambda_{l} \frac{\gmat{B}^{l}}{\Vert \gmat{B}^{l} \Vert_{2}}\ ,
\end{eqnarray}
where the last two equation are true for $\Vert \gmat{B}^{l} \Vert_{2} \neq 0$. In this
case the matrix derivative of the cost function is:
\begin{eqnarray}
  \frac{\delta}{\delta \gmat{B}^{l}} \mathcal{C}(\gmat{B}) &=& 2\gmat{B}^{l} - 2\gmat{\Gamma}^{l} + \
  \lambda_{l} \frac{\gmat{B}^{l}}{\Vert \gmat{B}^{l} \Vert_{2}} + \
  \lambda \Sign{\gmat{B}^{l}}\ ,
\end{eqnarray}
minimizing the cost function give the equation:
\begin{eqnarray}
  \label{eq:grad}
  \gmat{B}^{l}\bigg( 2 + \frac{\lambda_{l}}{\Vert \gmat{B}^{l} \Vert_{2}}\bigg) &=& 2\gmat{\Gamma}^{l} - \lambda \Sign{\gmat{B}^{l}}\ .
\end{eqnarray}

This equation is only valid for $\Vert \gmat{B}^{l} \Vert_{2} \neq 0$,
otherwise we use the sub-gradient
method\cite[]{Friedman:10} to minimize the cost function:
\begin{eqnarray}
  \label{eq:subgrad}
  \frac{\delta}{\delta \gmat{B}^{l}} \mathcal{C}(\gmat{B}) &=& 2\gmat{B} - 2\gmat{\Gamma} +  \lambda \gmat{T}^{l} + \
  \lambda_{l}\vec{s}_{l} = 0 \,
\end{eqnarray}
subject to constraints:
\begin{eqnarray}
  \Vert \vec{s}_{l} \Vert_{2} < 1\cr
  t_{li} \in [-1,1]\ .
\end{eqnarray}
The logical twist here is that we do not know {\it a priori} what
$\Vert \gmat{B}^{l} \Vert_{2}$ ought to be. The strategy then is to look at
each row of the parameter matrix $\gmat{B}$ and determine if a solution to
\eqref{eq:subgrad} exists subject to the constraints. Where we do find
feasible solution to this inequality, we reason that the rows of
$\gmat{B}^{l}$ ought to be zero.

At each iteration, this reasoning leads to a two step process for
thresholding $\gmat{B}^{l}$, the first conforms to a sparse representation of
a layered Earth, the second requires a sparse representation in the
basis of polynomials we use to fit the AVO response. The solution
is considered to have converged when the change in log-likelihood
Eq.~\eqref{eq:master} is less than some threshold.

\subsection{Defining active sets}

Suppose $\gmat{B}^{l} = \vec{0}$, then
\begin{eqnarray}
  \lambda \gmat{T}^{l} + \lambda \vec{s}_{l} = 2\gmat{\Gamma}^{l}\ ,
\end{eqnarray}
the requirement that $\Vert \vec{s}_{l} \Vert_{2} < 1$ implies that:
\begin{eqnarray}
  \label{eq:thresh_1}
  \Vert \gmat{\Gamma}^{l} - \frac{\lambda}{2} \gmat{T}^{l}\Vert_{2} \leq \frac{\lambda_{l}}{2}\ .
\end{eqnarray}
If there exists any $\gmat{T}^{l}$ that satisfies this inequality, then we can conclude
that our supposition was correct. We can write the norm of this equation in
index notation:
\begin{eqnarray}
  \Vert \gmat{\Gamma}^{l} - \frac{\lambda}{2} \gmat{T}^{l}\Vert_{2} &=&
  \sqrt{\sum_{q=1}^{r}(\gamma_{lq} - \frac{\lambda}{2} t_{lq})^{2}}\ ,
\end{eqnarray}
the minimum of the norm can be found by minimizing
$(\gamma_{lq} - \frac{\lambda}{2} t_{lq})^{2}$ element by element. In which case,
if $\vert \gamma_{lq} \vert > \frac{\lambda}{2}$, then at the minima the element
$t_{lq} = \frac{2}{\lambda} \Sign{\gamma_{lq}}$, otherwise 
$t_{lq} = \frac{2}{\lambda} \gamma_{lq}$. To determine if there exists a solution
to the inequality \eqref{eq:thresh_1}, it is sufficient to determine whether:
\begin{eqnarray}
  \argmin_{T^{l}} \Vert \gmat{\Gamma}^{l} - \frac{\lambda}{2} \gmat{T}^{l}\Vert_{2} \leq \frac{\lambda_{l}}{2}\ ,
\end{eqnarray}
if this inequality can be satisfied, then $\gmat{B}^{l}$ is thresholded to zero. In the parlance
of the group lasso algorithm\cite[]{Friedman:10}, the non-zero rows of $\gmat{B}$ define the
active sets.

The consequence of this first thresholding operation is to define a discrete subset of
time samples where the reflectivity is allowed to be non-zero. We can think of the penalty
$\mathcal{P}_{1}(\gmat{B})$, as enforcing sparsity in the layered Earth.

A corollary to this argument is where $\Vert \gmat{B}^{l} \Vert_{2} \neq 0$. In this case
\begin{eqnarray}
  \label{eq:corol}
  \Vert \gmat{\Gamma}^{l} - \frac{\lambda}{2} \Sign{\gmat{B}^{l}}\Vert_{2} > \frac{\lambda_{l}}{2}\ ,
\end{eqnarray}
this identity holds for each of the active sets. The proof is by
contradiction; it is true that $\Sign{\gmat{B}^{l}} \in [-1,1]$. Suppose then
that the inequality did not hold. That would imply that there existed
$T^{l} = \Sign{\gmat{B}^{l}}$ on the interval $t_{li} \in [-1,1]$, such that
Eq.~\eqref{eq:thresh_1} was satisfied.  In which case $\gmat{B}^{l}$ was not
an active set.

\subsection{Soft thresholding}

Having defined the active sets, we can then proceed to solve Eq.~\eqref{eq:grad}.
First solving for $\Vert \gmat{B}^{l} \Vert_{2}$, we take the norm of both sides of the
equation, and solve:
\begin{eqnarray}
  \Vert \gmat{B}^{l} \Vert_{2} = \Vert \gmat{\Gamma}^{l} - \frac{\lambda}{2} \Sign{\gmat{B}^{l}}\Vert_{2} - \frac{\lambda_{l}}{2}\ .
\end{eqnarray}
The corollary of the previous section guarantees that the right hand side of this equation
is positive.

Substituting the solution of the norm back into Eq.~\eqref{eq:grad} (and doing some algebra) we find
the solution:
\begin{eqnarray}
  \gmat{B}^{l} &=& \bigg( 1 - \frac{\lambda_{l}}{2\Vert \gmat{\Gamma}^{l} -
    \frac{\lambda}{2}\Sign{\gmat{B}^{l}}\Vert_{2}} \bigg)(\gmat{\Gamma}^{l} - \frac{\lambda}{2}\Sign{\gmat{B}^{l}})
\end{eqnarray}
Again, by the corollary of the previous section,
\begin{eqnarray}
  1 - \frac{\lambda_{l}}{2\Vert \gmat{\Gamma}^{l} -
    \frac{\lambda}{2}\Sign{\gmat{B}^{l}}\Vert_{2}}  > 0\ ,
\end{eqnarray}
in which case:
\begin{eqnarray}
\Sign{\gmat{B}^{l}} &=& \Sign{\gmat{\Gamma}^{l} - \frac{\lambda}{2}\Sign{\gmat{B}^{l}}}\ .
\end{eqnarray}
We need to search element by element to find a solution that makes sense, this is our second
``soft'' thresholding step:
\begin{equation}
  b_{li}=\left\{
  \begin{array}{@{}ll@{}}
    \bigg( 1 - \frac{\lambda_{l}}{2\Vert \gmat{\Gamma}^{l} -
      \frac{\lambda}{2}\Sign{\gmat{B}^{l}}\Vert_{2}} \bigg)  
    (\gamma_{li} - \frac{\lambda}{2} \Sign{\gamma_{li}} ), & \text{if}\ \vert {\gamma_{li}}\vert > \frac{\lambda}{2} \\
    0, & {\rm otherwise}
  \end{array}\right.
\end{equation} 

\subsection{Optimization algorithm}

Following Daubechies\cite[]{Daub:04}, the optimization algorithm is detailed in
Algorithm~\ref{algo}.
\IncMargin{1em}
\begin{algorithm}
   \KwData{$\gmat{Y}$, $\gmat{A}$ and $\gmat{W}$}
   \KwResult{$\gmat{B} = \tilde{\gmat{B}}$}
   \SetKwData{active}{\ Active Sets}\
   \SetKwData{Btilde}{$\tilde{\gmat{B}}$}\   
   \Btilde$\leftarrow \gmat{Y}\gmat{A}$\;
   \While{Not converged}{
     Compute $\gmat{\Gamma} = (\gmat{W}^{T}\gmat{Y}\gmat{A} - \gmat{W}^{T}\gmat{W} \gmat{B} + \tilde{\gmat{B}})$\;
     \tcc{First threshold step}{}
     \lForEach{row $l$}
              {
              \begin{equation}                
                t_{li} = \left\{\begin{array}{@{}ll@{}}
                \frac{2}{\lambda}\Sign{\gamma_{li}}, & \vert \gamma_{li}\vert > \frac{\lambda}{2}\\
                \frac{2}{\lambda} \gamma_{li}, & {\rm otherwise}
                \end{array}\right.
              \end{equation}
              \lIf{$\Vert \gmat{\Gamma}^{l} - \frac{\lambda}{2}\gmat{T}^{l}\Vert_{2} \leq \frac{\lambda_{l}}{2}$}
                  {
                    $\gmat{B}^{l} = \vec{0}$
                  }
                  \lElse{\active$\leftarrow l$}
              }
     \tcc{Second threshold step}{}
              \lForEach{$l \in $\active}
                       {
                         \begin{equation}
                           b_{li}=\left\{
                           \begin{array}{@{}ll@{}}
                             \bigg( 1 - \frac{\lambda_{l}}{2\Vert \gmat{\Gamma}^{l} -
                               \frac{\lambda}{2}\Sign{\gmat{B}^{l}}\Vert_{2}} \bigg)  
                             (\gamma_{li} - \frac{\lambda}{2} \Sign{\gamma_{li}} ), \
                             & \text{if}\ \vert {\gamma_{li}}\vert > \frac{\lambda}{2} \\
                             0, & {\rm otherwise}
                           \end{array}\right.
                         \end{equation} 
                         
                         }
              \Btilde$\leftarrow \gmat{B}$\;
   }
\caption{Optimization with dual sparsity constraints.}\label{algo}
\end{algorithm}\DecMargin{1em}

\subsection{Convergence}

The principle metric of convergence is that the function:
\begin{eqnarray}
  L(\gmat{B}) &=& C_{0}(\gmat{B}) + P_{0}(\gmat{B}) + P_{1}(\gmat{B}) \ ,
\end{eqnarray}
should be monotonically decreasing as we iterate, that is for the $n^{th}$ iteration,
$L(\gmat{B}^{n}) < L(\gmat{B}^{n-1})$. In fact, this inequality must always hold, and therefore is
also good test for correctness, our implementation monitors this criteria as the
algorithm iterates.

We found that the cost function $L(\gmat{B})$ monotonically decreased at each
iteration for both the synthetic and applied example to within
machine precision.

\bibliography{bib}
\bibliographystyle{seg}

\end{document}